\documentclass[journal,12pt,onecolumn]{IEEEtran}
\IEEEoverridecommandlockouts
\usepackage[english]{babel}                             
\makeatletter
\adddialect\l@ENGLISH\l@english
\makeatother
\usepackage[T1]{fontenc}
\usepackage{pifont}
\newcommand{\cmark}{\ding{51}}%
\newcommand{\xmark}{\ding{55}}%

\usepackage{mathtools}
\usepackage{amsmath, amssymb, amsthm}
\usepackage{bm}
\usepackage[inline]{enumitem}

\usepackage[switch]{lineno}
\usepackage[named]{algo}
\usepackage[noend]{algpseudocode}
\usepackage{algorithm}

\usepackage{float}
\usepackage{stfloats}
\usepackage[section]{placeins}
\usepackage{balance}
\usepackage{etoolbox}
\usepackage{xcolor}

\usepackage[pdftex]{graphicx}
\usepackage[none]{hyphenat}

\usepackage{lipsum}
\usepackage{graphicx}
\ifCLASSOPTIONcompsoc
    \usepackage[caption=false, font=normalsize, labelfont=sf, textfont=sf]{subfig}
\else
\usepackage[caption=false, font=footnotesize]{subfig}
\fi

\newtheorem{theorem}{Theorem}

\newtheorem{proposition}[theorem]{Proposition}

\newtheorem{claim}{Claim}
\newtheorem{remark}{Remark}

\newtheoremstyle{colon}%
{}
{}
{\rm}
{}
{\itshape}
{:}
{ }
{\thmname{#1}\thmnumber{ \itshape#2}\thmnote{ (#3)}}

\theoremstyle{colon}

\newcommand{\dddag}{%
  \mathbin{\vbox{\offinterlineskip\ialign{%
    \hfil##\hfil\cr
    \small{$\dagger$}\cr
    \noalign{\kern-0.5ex}
    \small{$\ddagger$}\cr
}}}}

\graphicspath{{Figures/}}

\begin{document}
\setlength{\abovedisplayskip}{3pt}
\setlength{\belowdisplayskip}{3pt}

\linespread{1.5}
\title{\Large Clustered Cell-Free Multi-User MIMO Systems with Rate-Splitting \vspace{-0.5em} }

\author{\normalsize Andre R. Flores, Rodrigo C. de Lamare, and Kumar Vijay Mishra 

\thanks{A. R. F. and R. C. d. L. are with the Pontifical Catholic University of Rio de Janeiro, Brazil. E-mail: \{andre.flores, delamare\}@cetuc.puc-rio.br. K. V. M. is with the United States DEVCOM Army Research Laboratory, Adelphi MD, 20783 USA. E-mail: kvm@ieee.org. The conference precursor of this paper appeared in the 2022 IEEE International Conference on Communications (ICC).}}




%


\maketitle

\begin{abstract}
In this paper, we address two crucial challenges in the  design of cell-free (CF) systems: degradation in the performance of CF systems by imperfect channel state information at the transmitter (CSIT) and high computational/signaling loads arising from the increasing number of distributed antennas and parameters to be exchanged. To mitigate the effects of imperfect CSIT, we employ rate-splitting (RS) multiple-access, which separates the messages into common and private streams. Unlike prior works, we present a clustered CF multi-user multiple-antenna framework with RS, which groups the transmit antennas in several clusters to reduce the computational and signaling loads. The proposed RS-CF system employs one common stream per cluster to exploit the network diversity. Furthermore, we propose new cluster-based linear precoders for this framework. We then devise a power allocation strategy for the common and private streams within clusters and derive closed-form expressions for the sum-rate performance of the proposed cluster-based RS-CF system. Numerical results show that the proposed clustered RS-CF system and algorithms outperform existing approaches. 
\end{abstract}


\setlength{\abovedisplayskip}{3pt}
\setlength{\belowdisplayskip}{3pt}

%
\IEEEpeerreviewmaketitle

\vspace{-1.25em}
\section{Introduction}
The current infrastructure of wireless communications systems relies on base stations (BSs), which are deployed over the area of interest and provide services to multiple users. However, networks employing standard centralized BSs do not meet the increasing requirements that future services, such as virtual reality and the internet of things demand. Indeed, future applications require higher data rates, reliable connectivity, better quality-of-service and lower latency than the previous standards \cite{Harsh2021}. Although the densification of BSs would help to provide future services, it results in stronger multi-user interference (MUI) \cite{Tse2005} that is more difficult to handle.


In this context, cell-free (CF) multiple-input multiple-output (MIMO) systems are viewed as a potential technology to satisfy the demands of future wireless networks \cite{Elhoushy2021}. Instead of deploying multiple centralized BSs, CF MIMO systems employ multiple distributed access points (APs) that are connected to a central processing unit (CPU). These APs serve a small group of users that are geographically distributed. It was shown in \cite{Ngo2017,Nguyen2017} that CF MIMO systems achieve
higher energy efficiency (EE) and throughput per user than conventional systems employing BSs. As a result, CF deployment has garnered significant research interest.

\vspace{-1em}
\subsection{Prior Art}
\vspace{-0.5em}
The receivers in a CF network experience MUI because of the simultaneous transmission to multiple users using the same time-frequency resources. In general, precoding techniques are implemented in the CF downlink to mitigate MUI. A popular low-complexity precoding technique for CF is conjugate beamforming (CB) or matched filter (MF) precoder \cite{Ngo2017}. To further enhance the performance of the MF precoder in the high SNR regime, the zero-forcing (ZF) precoder was investigated in \cite{Nguyen2017} for CF systems. In \cite{Nayebi2017}, the design of CB and ZF precoders was combined with power allocation to provide rate fairness between the users. The literature suggests that the minimum-mean square error (MMSE) precoder outperforms the CB and ZF precoding techniques. Therefore, iterative minimum mean square error (MMSE) precoders for CF, where the precoders and power allocation are updated at each iteration were proposed in \cite{Palhares2021}. Most of these works employ network-wide receivers or precoders. However, this approach is not feasible due to the increasing signaling and computational loads. To save resources and decrease the power consumption of CF systems, \cite{Palhares2021a} proposed AP selection. Indeed, several works \cite{Buzzi2017,Buzzi2020,Flores2022b} suggest curtailing the number of APs and users that are jointly processed. Recently, scalable MMSE precoders and combiners have also been put forward to facilitate the deployment of CF systems \cite{Bjoernson2020}. In \cite{Lozano2021}, the number of APs serving each user was reduced by forming subsets and employing the regularized ZF precoder.

Precoder design in the above-mentioned works assumes that perfect channel state information at the transmitter (CSIT) is available. The CSIT is obtained by employing pilot sequences along with the reciprocity properties in systems using time-division duplex (TDD) and feedback channels in frequency-division duplex (FDD) techniques \cite{Vu2007}. Nevertheless, in practice, considering perfect CSIT is an unrealistic assumption because several sources of errors degrade the quality of channel estimates. For example, pilot contamination from noise adversely affects the estimation procedure. Moreover, time-varying dynamic channels require constant CSIT updates. Consequently, the transmitter has only partial or imperfect CSIT. As a result, 
the precoder can no longer deal with the MUI as expected. 
The residual MUI at the receiver degrades heavily the system performance because it scales with the transmit power \cite{Tse2005}. {Therefore, 
new transmission techniques that take into account imperfect CSIT are urgently needed.}

Rate-splitting (RS) \cite{Clerckx2016,Mao2022} has emerged as an approach that can address imperfect CSIT more effectively than conventional schemes. Tracing back its origins to \cite{Han1981}, RS deals with interference channels \cite{Carleial1978}, where independent transmitters sent information to independent receivers \cite{Haghi2021}. RS was extended in \cite{Yang2013} to the broadcast channel of MIMO systems \cite{mmimo,wence}, where it was shown to provide gains in terms of degrees-of-freedom (DoFs) over conventional multi-user MIMO under imperfect CSIT. In particular, \cite{Piovano2017} demonstrated that RS achieves the optimal DoF region under imperfect CSIT. Moreover, RS has been proven to be robust against channel imperfections and other degrading effects of user mobility\cite{Dizdar2021}.
RS transmissions split the message of one or several users into common and private messages. The common message must be decoded by all the users while the private messages are decoded only by their corresponding users. Unlike conventional schemes, such as spatial division multiplexing (SDMA) and the power-domain non-orthogonal multiple access (NOMA), RS is capable of adjusting the content and the power of the common message, offering robustness against imperfect CSIT by controlling the amount of interference to be decoded or treated as noise. Indeed, RS outperforms the conventional SDMA and NOMA, by achieving higher rates than both of them \cite{Mao2018}. Surprisingly, RS has shown robustness 
against even dirty paper coding \cite{Mao2020}. Interestingly, it constitutes a generalized framework that has other transmission techniques such as SDMA, NOMA, and multicasting as its special cases \cite{Clerckx2020,Naser2020,Jaafar2020}.

Since then, several deployments and performance metrics involving RS have been studied with linear \cite{siprec,gbd,wlbd,lrcc} and non linear precoders \cite{mbthp,rmbthp,bbprec,rsthp}. In particular, \cite{JoudehClerckx2016,Hao2015} investigated the sum-rate maximization in multiple-input single-output (MISO) networks employing RS along with linear precoders. Furthermore, overloaded multigroup multicasting scenarios in \cite{Joudeh2017} considered both the sum-rate maximization criterion and max-min fairness approach. In \cite{Lu2018}, RS was used in a system with random vector quantization feedback. RS with common stream combining techniques \cite{Flores2020,rsthp} exploited multiple antennas at the receiver and improved the overall sum-rate performance. The single-cell RS study in \cite{Li2020} provided algorithms to reduce the number of streams and perform successive decoding for a large number of users. An RS approach suited for massive MIMO environments known as hierarchical-rate-splitting (HRS) was proposed in \cite{Dai2016}. However, the performance of massive MIMO architectures drops substantially due to pilot contamination. To address this problem, a robust implementation to mitigate the effects of pilot contamination in RSMA massive MIMO systems has been proposed in \cite{Mishra2022}. RS-CF architectures have been reported in \cite{Flores2022,Mishra2022}, where small networks have been considered.
{In this work, we focus on the hitherto unexplored application of RS in cluster-based CF systems. Table \ref{tabela: Literature Review} summarizes the differences between the proposed system and prior works.}

\begin{table}[t]
\centering 

\caption{Comparison with the state of the art}
\vspace{-1em}
\small
\begin{tabular}{c|c c c c c c}
\hline 
q.v. & \multicolumn{3}{c}{System Model} & Precoder & Signaling Load & Performance Metric \\ [0.5ex]
 & RS & CF & AP Clusters & &  \\
\hline
\hline 
\cite{Mishra2022}& \cmark & \cmark & \xmark & Non-clustered & Moderate  & Sum-rate  \\
\cite{Mishra2022b}& \cmark & \xmark & \xmark & Non-clustered & High & Sum-rate, product of SINR, fairness  \\
\cite{Park2022} & \cmark & \xmark & \xmark & Non-clustered & High & Sum-rate   \\
\cite{Dai2016} & \cmark & \xmark &\xmark & Non-clustered & High &Asymptotic sum-rate  \\
\cite{Chen2021}& \xmark & \cmark &\cmark &Clustered & Low & CDF \\
This paper &\cmark & \cmark & \cmark &Clustered & Low &Sum-rate \\
 [1ex] 
\hline 
\end{tabular}\normalsize
\label{tabela: Literature Review} 

 \vspace{-2em}
\end{table}

 \vspace{-1em}
\subsection{Contributions}
 \vspace{-0.5em}
Preliminary results of this work appeared in our conference publication \cite{Flores2022}. In this work, we propose a cluster-based CF architecture that employs an RS scheme to transmit the information to the users and mitigate the effects of imperfect CSIT. Different from {\cite{Flores2022} and \cite{Mishra2022}, the proposed RS-CF scheme separates the users into several disjoint clusters to transmit multiple common messages. The clusters are formed based on the large-scale fading coefficients following a user-centric approach. In particular, one common message per cluster is sent, generating a common stream with a power allocation for the cluster. The rationale behind forming disjoint clusters and sending one common stream per cluster is that the contribution to the received signal quality of far away APs is small as compared to other APs. Thus, the interference caused by the common messages of other clusters is small. In this sense, each common message handles imperfect CSIT at each cluster, and, therefore, the conventional CF structure obtains benefits from the RS transmission scheme.

On the other hand, RS also benefits from the CF architecture. The performance attained by the common stream in conventional RS systems is limited by the worst user so that all users decode the common message. Therefore, a user which is poorly served degrades heavily the overall performance of the system. By distributing the APs, CF provides better conditions for the channels of all users, enhancing the performance of the common stream. Moreover, the proposed cluster-based RS-CF architecture restricts the number of users per common stream, providing extra gains in terms of the common rate. {All users in a cluster must decode first the common message intended for its cluster while treating other common messages as noise.} Once the common message is decoded, the receivers decode their private messages.

Power allocation is a crucial step in RS systems. In conventional RS transmissions, the rate performance is degraded heavily when the power allocated to the common stream is not done properly. The proposed architecture transmits multiple common streams to different clusters, which results in different power allocations for each common stream. In this sense, power allocation becomes more challenging as compared to conventional RS architectures. To keep the computational cost low, we resort to a simplified exhaustive search algorithm to carry out power allocation, which limits the number of parameters that must be found to avoid an exponential cost. The proposed RS-CF approach obtains gains in terms of ergodic sum-rate (ESR) over standard CF and RS architectures.

We also develop cluster-based precoders, which include one common precoder obtained per cluster based on a singular value decomposition (SVD) over the channel matrix. In order to send the private symbols, we devise cluster-based linear MF, ZF, and MMSE precoders, which are 
{implementation-friendly} and cost-effective. In general, cluster-based precoders are implemented to address scalability issues. We present an analysis to show that the proposed clustered RS-CF system is scalable. We also derive closed-form expressions for the signal-to-interference-plus-noise (SINR) and the ESR of this new clustered RS-CF. In contrast to prior works, the clustered RS strategy provides a consistent gain from the multiple common messages. The clustered CF approach further improves RS by controlling the common message each user should decode. Simulation results show that combining CF with RS yields a consistently increasing ESR even in the high signal-to-noise ratio (SNR) regime, robustness against imperfect CSIT, and significant sum-rate performance gains over standard MU-MIMO networks.

{The proposed RS-CF architecture, therefore, leverages the benefits of both RS and clustered-CF to improve both systems}. The contributions of this work are summarized as follows:
\begin{itemize}
    \item A clustered RS-CF architecture that is robust against imperfect CSIT and provides excellent sum-rate performance;
    \item Cluster-based efficient linear MF, ZF, and MMSE precoders that have low signaling load;
    \item A clustering algorithm to form disjoint groups of users and APs.
    \item A joint power allocation algorithm to adjust the power of the common streams across clusters.
    \item Analyses of the sum-rate, computational cost, and signaling load of the proposed RS-CF and precoders;
    \item A simulation study of both the proposed RS-CF and the existing centralized and CF architectures, including the respective precoders.
\end{itemize}

 \vspace{-1.5em}
\subsection{Organization}
The rest of this paper is organized as follows. In the next section, 
we describe the system model of a CF MIMO system. We introduce the proposed clustered RS-CF in Section~\ref{sec:rs}. The design of the AP selection method and cluster precoders is detailed in Section~\ref{sec:ap}.  In Section~\ref{sec:alloc}, we present the joint power allocation of the common streams across clusters. The analyses of the sum-rate performance, computational complexity, and signaling load are developed in Section~\ref{sec:perf}. Numerical examples are shown and discussed in Section~\ref{sec:numexp}, and concluding remarks are drawn in Section~\ref{sec:summ}.

Throughout the paper, we reserve bold lowercase, bold uppercase, and calligraphic letters for the vectors, matrices, and sets respectively; $\textrm{Tr}(\cdot)$ and $\mathbb{E}\left[\cdot\right]$ represent trace and statistical expectation operators, respectively; the notations $(\cdot)^{\text{T}}$, $(\cdot)^H$, $(\cdot)^{*}$, $\lVert\cdot\rVert$ and $|\cdot|$ denote the transpose, Hermitian, complex conjugate, Euclidean norm, and magnitude, respectively; the operator $\odot$ denotes the Hadamard product. An $N \times K$ matrix with column vectors $\mathbf{a}_1$, $\cdots$, $\mathbf{a}_K$, each of length $N$, is $\mathbf{A} = \left[\mathbf{a}_1,\cdots,\mathbf{a}_K\right]$. A diagonal matrix with the elements of vector $\mathbf{v}$ in the main diagonal is $\textrm{diag}(\mathbf{v})$. The operator $\Re\left\lbrace \cdot \right\rbrace$ retains the real part of a complex argument. The notation $a \sim \mathcal{CN}(0,\sigma_a^2)$ denotes circularly symmetric complex Gaussian random variable $a$ with zero mean and variance $\sigma_a^2$. We denote an $N\times 1$ vector of all ones by $\mathbf{1}_N$.

\vspace{-1.5em}
\section{System Model}
\label{sec:sysmod}



Consider the downlink of a CF wireless network, wherein the users
are geographically distributed. In this work, we use the terms user
and user equipment (UE) interchangeably to refer to a wireless
device that communicates with the APs. Assume $M$ randomly
distributed single-antenna APs cover the geographical area of
interest. The distributed APs serve $K$ users in an under-loaded
regime i.e., $M> K$, where each user is equipped with a single
antenna. This differs from a standard RS cellular network that
assigns the users to specific BSs. All antennas are connected to an
RS-based central processing unit (CPU) that may be located in a
cloud.

The information is split, encoded, and modulated before the
transmission. The APs send the transmit vector $\mathbf{x}\in
\mathbb{C}^{M}$ to the users. The system follows a transmit power
constraint, i.e.,
$\mathbb{E}\left[\lVert\mathbf{x}\rVert^2\right]\leq P_t$. We assume
a flat-fading channel model where the channel coefficient between
the $m$-th AP and the $k$-th user is
\begin{equation}
\label{eq:g}
    g_{m,k}=\sqrt{\zeta_{m,k}}h_{m,k},
\end{equation}
where $\zeta_{m,k}$ are the large-scale fading coefficients that
incorporate the path loss and shadowing effects and $h_{m,k}$
represent the small-scale fading coefficients that independently and
identically distributed (i.i.d.) random variables following the
distribution $\mathcal{CN}\left(0,1\right)$.
The coherence time $\tau_c$ is the interval during which the channel impulse response and
the small-scale fading coefficients do not change. 
The channels between all APs and users are given by
 $\mathbf{G}=\left[\boldsymbol{g}_1,\boldsymbol{g}_2,\cdots,\boldsymbol{g}_K\right]\in \mathbb{C}^{M\times K}$. 

The system employs the TDD protocol and, therefore, the channel is
estimated by exploiting the channel reciprocity property and pilot
training. First, all users simultaneously and synchronously transmit
the pilot sequences
$\boldsymbol{\pi}_{1},\cdots,\boldsymbol{\pi}_{K}\in\mathbb{C}^{\tau}$ to acquire the CSI at the transmitter.
After receiving the pilots, we compute the channel estimate as
$\mathbf{\hat{G}}=\left[\boldsymbol{\hat{g}}_1,\boldsymbol{\hat{g}}_2,\cdots,\boldsymbol{\hat{g}}_K\right]\in\mathbb{C}^{M\times K}$, 
wherein $\{\boldsymbol{\hat{g}}_k\}_{k=1}^K$ denote the columns. The
$(m,k)$-th element of the matrix $\mathbf{\hat{G}}$ is the channel
estimate between the $m$-th AP and the $k$-th user, i.e.,
\begin{equation}
    \hat{g}_{m,k}=\sqrt{\zeta_{m,k}}\left(\sqrt{1-\sigma_e^2}h_{m,k}+\sigma_e\tilde{h}_{m,k}\right),\label{channel coefficient estimate and channel relation}
\end{equation}
where $\tilde{h}_{m,k}$ models the error of the channel estimate and
follows a complex Gaussian distribution with zero mean and unit
variance, and $\sigma_e$ denotes the quality of the channel
estimate.

The CSI is then sent to the CPU 
for precoding. The channel estimation errors are represented by the ${M\times K}$ complex matrix
$\mathbf{\tilde{G}}=\left[\boldsymbol{\tilde{g}}_1,\boldsymbol{\tilde{g}}_2,\cdots,\boldsymbol{\tilde{g}}_K\right]$, 
whose $(m,k)$-th element
\begin{align}
\label{eq:gtilde}
\tilde{g}_{m,k}=\sigma_e\sqrt{\zeta_{m,k}}\tilde{h}_{m,k},
\end{align}
represents the error affecting the channel estimate between the
$m$-th AP and the $k$-th user. Then, the estimates of the channels
and their associated errors are, respectively,
$\hat{g}_{m,k}\sim\mathcal{CN}\left(0,\zeta_{m,k}\right),$ and
$\tilde{g}_{m,k}\sim\mathcal{CN}\left(0,\sigma^2_{e}\zeta_{m,k}\right).$
The received signal vector $\mathbf{y}^{\left(\text{CF}\right)} =
[y_1^{\left(\text{CF}\right)}, \cdots,
y_K^{\left(\text{CF}\right)}]^T$ is
\begin{equation}
    \mathbf{y}^{\left(\text{CF}\right)}=\mathbf{G}^{\text{T}}\mathbf{x}+\mathbf{w},
\end{equation}
where the vector
$\mathbf{w}=\left[w_1,w_2,\cdots,w_K\right]^{\text{T}}\sim\mathcal{CN}\left(\mathbf{0},\sigma^2_w\mathbf{I}\right)$
is the additive white Gaussian noise (AWGN) at the receiver.

\begin{figure}[t]
\begin{center}
\includegraphics[width=0.35\columnwidth]{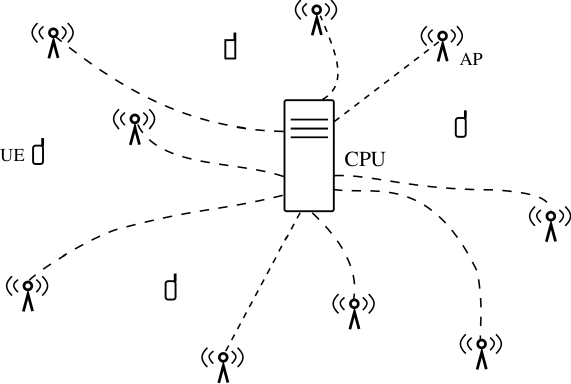}
\vspace{-1.25em} \caption{Illustration of the CF MIMO system with
randomly distributed APs and UEs. The CPU coordinates the APs.
\vspace{-1.5em}} \label{Fig1} \vspace{-1.0em}
\end{center}
\end{figure}


\begin{figure}
    \centering
  \subfloat[\label{2a}]{%
       \includegraphics[width=0.4\linewidth]{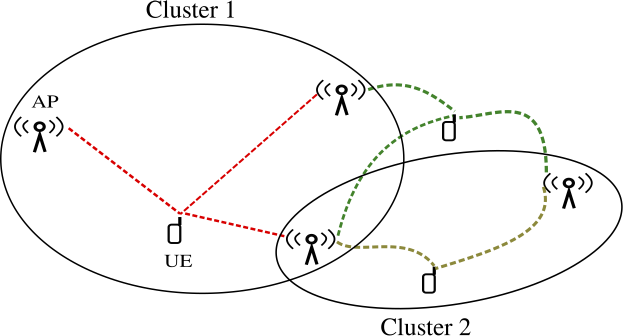}}
    \hfill
  \subfloat[\label{1b}]{%
        \includegraphics[width=0.4\linewidth]{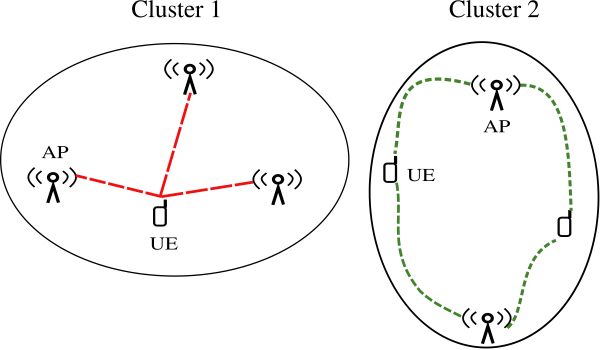}}
\vspace{-1.1em}
  \caption{(a) A cluster-based CF network, where APs may be shared between Clusters 1 and 2. (b) A non-overlapping cluster-based CF network with two clusters.} \vspace{-1.5em}
  \label{Fig2}
\end{figure}


\section{RS for Cell-Free Architectures}

\label{sec:rs}

The computational cost of a CF network-wide precoder is extremely
high and unsuitable for practical systems. To address this problem,
cluster-based precoders can be employed, which reduce significantly
the amount of signaling required. Moreover, a cluster-based approach
can also simplify the computation of the precoder, saving even more
resources. The clusters are constructed based on the large-scale
fading coefficients of the channel. The motivation for this
structure is that only a few APs have a major contribution to the
intended receive signal. Other APs, which may be far away or
experiencing bad channel conditions, {barely contribute} to the
intended signal and could be discarded. Fig. \ref{Fig2} shows
cluster-based approaches for CF networks.

Different from conventional RS schemes, a cluster-based approach should take into account the distribution of the network. Therefore, we propose a clustered RS-based CF structure where multiple common streams are sent. The APs are separated into disjoint clusters with $N_c$ denoting the number of clusters formed. A total of $N_c$ common streams are transmitted, each one associated with a different cluster. It follows that each AP transmits only one of the $N_c$ common messages to its cluster. Common messages from other clusters are considered noise. Then, at each cluster, the common message is decoded first. Once the common messages are decoded, each receiver decodes its private message. We consider that one private stream per user is transmitted, i.e., a total of $K$ private streams are sent over the channels.

Let us denote the cluster of APs that send the common symbol $s_{c_i}$ to a set $\mathcal{K}_i$ of users by $\mathcal{A}_i$. The clusters are assumed to be disjoint for simplicity, i.e., $\mathcal{A}_i\cap \mathcal{A}_j=\varnothing$, $\;\forall\; i,j=1,2,\cdots,N_c$. In other words, no AP may belong to two different clusters. Similarly,  $\mathcal{K}_i\cap\mathcal{K}_j=\varnothing, \;\forall\; i,j=1,2,\cdots,N_c$, i.e., all users receive only one common message. The clusters are separated based on the large-scale fading coefficient $\zeta_{m,k}$. The rationale behind the cluster-based transmission scheme is that only a few APs contribute to the most relevant part of the intended signal. For instance, consider a general cluster of APs denoted by $\mathcal{A}_i$. The remaining APs in $\mathcal{A}_j$ with $i\neq j$ experience unfavorable transmit conditions to users in $\mathcal{K}_i$, requiring extra power consumption and causing additional interference. Therefore, these APs are not appropriate for transmitting the common symbol $s_{c_i}$.

Forming clusters of APs leads to a reduced number of APs serving each user thereby decreasing the signaling load. Define the equivalent channel estimate as 
$\overline{\mathbf{G}}=\left[\overline{\boldsymbol{g}}_1,\overline{\boldsymbol{g}}_2,\cdots,\overline{\boldsymbol{g}}_K\right]\in \mathbb{C}^{M \times K}$. Considering that the $m$-th AP belongs to cluster $\mathcal{A}_i$, the $(m,k)$-th element of 
$\overline{\mathbf{G}}$ is 
\begin{equation}
    \overline{g}_{m,k}=\begin{cases}
\hat{g}_{m,k},&m\in \mathcal{A}_i, k\in \mathcal{K}_i\\
0, &\text{otherwise,}
\end{cases}
\label{sparse effective channel}
\end{equation}
It follows that the equivalent channel matrix is sparse and \eqref{sparse effective channel} can be used to compute the sparse precoders, thereby providing a reduction in the signaling load.

In our proposed RS cluster-based CF system, both common and private messages are encoded and modulated into a vector of symbols. The $N_c$ common symbols are {superposed} 
to the private symbols. Denote the vector of all symbols by  $\mathbf{s}^{\left(\text{RS}\right)}=\left[s_{c_1},s_{c_2},\cdots,s_{c_{N_c}},s_1,s_2,\cdots, s_K\right]^{\text{T}} \in \mathbb{C}^{K+N_c}$, where $s_{c_i}$ denotes the $i$-th common stream and $s_k$  stands for the $k$-th private symbol intended for user $k$. The symbols are mapped to the transmit antennas by employing a precoding matrix $\mathbf{P}^{\left(\text{RS}\right)}=\left[\mathbf{p}_{c_1},\mathbf{p}_{c_2},\cdots,\mathbf{p}_{c_{N_c}},\mathbf{p}_1,\mathbf{p}_2,\cdots, \mathbf{p}_K\right]\in\mathbb{C}^{M\times\left(K+N_c\right)}$, where the $M\times 1$ vectors $\mathbf{p}_{c_i}$ and $\mathbf{p}_k$ map the $i$-th common symbol and the $k$-th private symbol, respectively. Note that one common precoder per cluster is included to map each common message to its corresponding cluster. Power allocation is carried out using \textcolor{red}{the} matrix $\mathbf{A}^{\left(\text{RS}\right)}=\text{diag}\left([a_{c_1},\cdots,a_{c_{N_c}},a_1,\cdots, a_K]^{\text{T}}\right)   \in \mathbb{R}^{\left(K+N_c\right)\times\left(K+N_c\right)}$. 
The coefficient $a_{c_i}$ allocates a fraction of the available power $P_t$ to the $i$-th common stream, whereas the coefficient $a_k$ assigns the power to the $k$-th private symbol.

The APs send the transmit vector (that contains the private and common data) $\mathbf{x}^{\left(\text{RS}\right)}=\mathbf{P}^{\left(\textrm{RS}\right)}\mathbf{A}^{\left(\text{RS}\right)}\mathbf{s}^{\left(\textrm{RS}\right)} \in\mathbb{C}^{M}$  to the users. Then, the received signal vector $\mathbf{y} = [y_1, \cdots, y_K]^T$ of the proposed RS-CF architecture is
\begin{align}
    \mathbf{y} = &\mathbf{G}^{\text{T}}\mathbf{x}^{\left(\text{RS}\right)} + \mathbf{w}=\mathbf{G}^{\text{T}}\mathbf{P}^{\left(\textrm{RS}\right)}\mathbf{A}^{\left(\text{RS}\right)}\mathbf{s}^{\left(\textrm{RS}\right)} + \mathbf{w}.\label{RS-CF receive vector}
\end{align}
Expanding the terms of \eqref{RS-CF receive vector} by employing the relations established in \eqref{eq:g}-\eqref{eq:gtilde}, we get
\begin{align}   \mathbf{y}=&\varepsilon\left(\sum_{l=1}^{N_c}a_{c_l} s_{c_l} \mathbf{\hat{G}}^{\textrm{T}}\mathbf{p}_{c_l}-\sum_{j=1}^{N_c}a_{c_j} s_{c_j}\mathbf{\tilde{G}}^{\textrm{T}}\mathbf{p}_{c_j}+\sum_{q=1}^K a_q s_q \mathbf{\hat{G}}^{\textrm{T}}\mathbf{p}_q-\sum_{r=1}^K a_r s_r \mathbf{\tilde{G}}^{\textrm{T}}\mathbf{p}_r\right)+\mathbf{w},
\end{align}
where $\varepsilon=\frac{1}{\sqrt{1-\sigma^2_{e}}}$.
Then, the received signal at the $k$-th user is \par\noindent\small 
\begin{align}  
y_k=\varepsilon\left(\sum_{l=1}^{N_c}a_{c_l} s_{c_l}\hat{\boldsymbol{g}}_{k}^{\textrm{T}}\mathbf{p}_{c_l}-\sum_{j=1}^{N_c}a_{c_j} s_{c_j}\tilde{\boldsymbol{g}}_{k}^{\textrm{T}}\mathbf{p}_{c_j}
   +\sum_{q=1}^{K}a_q s_q\hat{\boldsymbol{g}}_{k}^{\textrm{T}}\mathbf{p}_{q}-\sum_{r=1}^{K}a_r s_r\tilde{\boldsymbol{g}}_{k}^{\textrm{T}}\mathbf{p}_{r}\right)+w_k,\label{received signal user k}
\end{align}\normalsize
Consider a cluster of APs in $\mathcal{A}_i$ that serve the user $k$ in $\mathcal{K}_i$. Rearranging the terms of \eqref{received signal user k} yields\par\noindent\small
\begin{align}
    y_k&=\varepsilon\left( \underbrace{a_{c_i}s_{c_i}\hat{\boldsymbol{g}}_{k}^{\textrm{T}}\mathbf{p}_{c_i}}_{T_c}+\underbrace{a_k s_k\hat{\boldsymbol{g}}_{k}^{\textrm{T}}\mathbf{p}_{k}}_{T_p} +\underbrace{\sum\limits_{\substack{l=1\\l\neq i}}^{N_c}a_{c_l} s_{c_l}\hat{\boldsymbol{g}}_{k}^{\textrm{T}}\mathbf{p}_{c_l}}_{I_{c}} +\underbrace{\sum_{\substack{q\in \mathcal{K}_i\\q\neq k}}a_q s_q\hat{\boldsymbol{g}}_{k}^{\textrm{T}}\mathbf{p}_{q}}_{I_{p,i}}+\underbrace{\sum_{t\notin \mathcal{K}_i}a_t s_t\hat{\boldsymbol{g}}_{k}^{\textrm{T}}\mathbf{p}_{t}}_{I_{p,o}}\right.\nonumber\\
    &\left.-\underbrace{\sum_{j=1}^{N_c}a_{c_j} s_{c_j}\tilde{\boldsymbol{g}}_{k}^{\textrm{T}}\mathbf{p}_{c_j}}_{I_{c,e}}-\underbrace{\sum_{r=1}^{K}a_r s_r\tilde{\boldsymbol{g}}_{k}^{\textrm{T}}\mathbf{p}_{r}}_{I_{p,e}} \right) +w_k,\label{received signal user k expanded}
\end{align}\normalsize
where $T_c$ denotes the common information that must be decoded, $T_p$ represents the private information of user $k$, $I_{p,i}$ stands for the interference caused by other users inside the cluster, 
$I_{p,o}$ denotes the interference produced by users outside the cluster, $I_c$ represents the interference generated from the common messages outside the cluster, $I_{c,e}$ is the residual common interference caused by the imperfect CSIT and $I_{p,e}$ quantifies the residual interference corresponding to the private information and caused by the imperfect CSIT.

Note that each user decodes only one common message. The receivers in $\mathcal{A}_i$ decode first the common symbol $s_{c_i}$. Then, successive interference cancellation (SIC) \cite{jidf,spa,mfsic,mbdf,bfidd,itic,1bitidd,did,dynovs,listmtc,dynmtc,jointmtc} is performed to subtract the common information from the received signal. We assume that perfect CSI is available at the receiver for simplicity \cite{Vu2007}. Removing the common symbol at the $k$-th user yields 
\par\noindent\small
\begin{align}
    y_k=&\varepsilon\left(a_k s_k\hat{\boldsymbol{g}}_{k}^{\textrm{T}}\mathbf{p}_{k} +\sum\limits_{\substack{j=1\\j\neq i}}^{N_c}a_{c_j} s_{c_j}\hat{\boldsymbol{g}}_{k}^{\textrm{T}}\mathbf{p}_{c_j}+\sum_{\substack{l\in \mathcal{K}_i\\l\neq k}}a_l s_l\hat{\boldsymbol{g}}_{k}^{\textrm{T}}\mathbf{p}_{l}+\sum_{q\notin \mathcal{K}_i}a_q s_q\hat{\boldsymbol{g}}_{k}^{\textrm{T}}\mathbf{p}_{q}\right.\nonumber\\
    &\left.-\sum_{\substack{j=1\\j\neq i}}^{N_c}a_{c_j} s_{c_j}\tilde{\boldsymbol{g}}_{k}^{\textrm{T}}\mathbf{p}_{c_j}-\sum_{r=1}^{K}a_r s_r\tilde{\boldsymbol{g}}_{k}^{\textrm{T}}\mathbf{p}_{r} \right)+w_k.
\end{align}\normalsize
Thereafter, the receiver recovers its private message.

From \eqref{received signal user k expanded}, the average power of the received signal at the $k$-th user is  \par\noindent\small
\begin{align}
    \mathbb{E}\left[\lvert y_k\rvert^2\right]=&\varepsilon^2\left(a_{c_i}^2\lvert\hat{\boldsymbol{g}}^{\text{T}}_k\mathbf{p}_{c_i}\rvert^2+a_k^2\lvert\hat{\boldsymbol{g}}_k^{\text{T}}\mathbf{p}_{k}\rvert^2+\sum\limits_{\substack{l=1\\l\neq i}}^{N_c}a_{c_l}^2\lvert\hat{\boldsymbol{g}}^{\text{T}}_k\mathbf{p}_{c_l}\rvert^2+\sum\limits_{\substack{q\in\mathcal{K}_i\\q\neq k}}a_q^2\lvert\hat{\boldsymbol{g}}^{\text{T}}_k\mathbf{p}_{q}\rvert^2+\sum\limits_{t\notin \mathcal{K}_i}a_t^2\lvert\hat{\boldsymbol{g}}^{\text{T}}_k\mathbf{p}_{t}\rvert^2+\sum\limits_{j=1}^{N_c}a_{c_j}^2\lvert\tilde{\boldsymbol{g}}^{\text{T}}_k\mathbf{p}_{c_j}\rvert^2\right.\nonumber\\
    &\left.+\sum\limits_{r=1}^K a_r^2\lvert\tilde{\boldsymbol{g}}^{\text{T}}_k\mathbf{p}_{r}\rvert^2-2\Re\left\{\sum_{j=1}^{N_c}a_{c_j}^2\left(\hat{\boldsymbol{g}}^{\text{T}}_k\mathbf{p}_{c_j}\right)^*\left(\tilde{\boldsymbol{g}}^{\text{T}}_k\mathbf{p}_{c_j}\right)\right\}-2\Re\left\{\sum_{r=1}^{K}a_{r}^2\left(\hat{\boldsymbol{g}}^{\text{T}}_k\mathbf{p}_{r}\right)^*\left(\tilde{\boldsymbol{g}}^{\text{T}}_k\mathbf{p}_{r}\right)\right\}\right)+\sigma_w^2.
\end{align}\normalsize
Then, the instantaneous SINR while decoding the $i$-th common symbol at the $k$-th user is
\par\noindent\small
\begin{align}
    \gamma_{c_i,k}&=\frac{a_{c_i}^2\lvert \boldsymbol{\hat{g}}_k^{\textrm{T}}\mathbf{p}_{c_i}\rvert^2}{d_{c_i,k}+\sum\limits_{\substack{j=1\\j\neq i}}^{N_c}a_{c_j}^2\lvert\boldsymbol{g}_k^{\text{T}}\mathbf{p}_{c_j}\rvert^2+\sum\limits_{r=1}^K a_r^2\lvert \boldsymbol{g}_k^{\textrm{T}}\mathbf{p}_r\rvert^2+\sigma_w^2/\varepsilon^2},\label{instantaneous SINR common rate}
\end{align}\normalsize
where $d_{c_i,k}=a_{c_i}^2\lvert{\tilde{\boldsymbol{g}}}^{\text{T}}_k\mathbf{p}_{c_i}\rvert^2-2a_{c_i}^2\Re\left\{\left({\hat{\boldsymbol{g}}}_k^{\text{T}}\mathbf{p}_{c_i}\right)^*\left({\tilde{\boldsymbol{g}}}_k^{\text{T}}\mathbf{p}_{c_i}\right)\right\}$ corresponds to the power loss arising from the error in the channel estimate. After applying SIC, the instantaneous SINR while decoding the private symbol at the $k$-th user, which belongs to $\mathcal{K}_i$, is \par\noindent\small
\begin{equation}
    \gamma_k=\frac{a_k^2\lvert\hat{\boldsymbol{g}}_k^{\textrm{T}}\mathbf{p}_k\rvert^2}{d_k+\sum\limits_{\substack{j=1\\j\neq i}}^{N_c}a_{c_j}^2\lvert\boldsymbol{g}_k^{\text{T}}\mathbf{p}_{c_j}\rvert^2+\sum\limits_{\substack{r=1\\r\neq k}}^K a_r^2\lvert\boldsymbol{g}^{\text{T}}_k\mathbf{p}_r\rvert^2+\sigma_w^2/\varepsilon^2},\label{instantaneous SINR private rate}
\end{equation}\normalsize
where $d_{k}=a_k^2\lvert\tilde{\boldsymbol{g}}^{\text{T}}_k\mathbf{p}_k\rvert^2-2a_k^2\Re\left\{\left({\hat{\boldsymbol{g}}}_k^{\text{T}}\mathbf{p}_k\right)^*\left(\tilde{\boldsymbol{g}}_k^{\text{T}}\mathbf{p}_k\right)\right\}$ corresponds to the power loss in the private stream arising from the imperfect CSIT.

Considering Gaussian signaling and employing \eqref{instantaneous SINR common rate}, we compute the instantaneous common rate (CR) of the proposed architecture at the $k$-th user in $\mathcal{K}_i$ as
\begin{equation}
R_{c_i,k}=\log_2\left(1+\gamma_{c_i,k}\right).
\label{instantaneous common rate}
\end{equation} Similarly, the instantaneous private rate (PR) at user $k$ is
\begin{equation}
R_k=\log_2\left(1+\gamma_k\right).
\label{instantaneous private rate}
\end{equation}Thus, the instantaneous sum-rate becomes
\begin{equation}
S_i=\sum_{j=1}^{N_c}\min_{k\in\mathcal{K}_j}R_{c_j,k}+\sum_{l=1}^{K} R_{l}.\label{instantaneous sum-rate}
\end{equation}Note that, in \eqref{instantaneous sum-rate}, we employ the minimum instantaneous CR found across the users in each cluster to guarantee that all users decode the common symbol successfully.

Unfortunately, the instantaneous rates computed directly with \eqref{instantaneous SINR common rate} and \eqref{instantaneous SINR private rate} are not achievable if we consider imperfect CSIT. To address this issue, we employ the average sum-rate (ASR) which averages out the effects of errors in the channel estimates. The ASR consists of two parts, the average CR and the average PR. The average CR at user $k$ is defined as the expected value of the instantaneous CR at user $k$ given a channel estimate, i.e., $\bar{R}_{c_j,k}=\mathbb{E}\left[R_{c_j,k}\left(\mathbf{G}\right)|\mathbf{\hat{G}}\right]$. Analogously, the average private rate at the $k$-th user is the expected value of the instantaneous PR given a channel estimate, i.e., $\bar{R}_{k}=\mathbb{E}\left[R_{k}\left(\mathbf{G}\right)|\mathbf{\hat{G}}\right]$. This yields the ASR as
\begin{equation} S_a=\sum_{j=1}^{N_c}\min_{k\in\mathcal{K}_j}\bar{R}_{c_j,k}+\sum_{l=1}^{K}\bar{R}_l. \label{system average sum rate}
\end{equation}

The performance of the system over a large number of channel realizations is measured by the ergodic sum-rate (ESR). Similar to ASR, the ESR is composed of the ergodic common rate (ECR) and the ergodic private rate (EPR). The ECR is the expected value of the ASR over multiple channel realizations, i.e., $R_{e,c}=\sum_{j=1}^{N_c}\min_{k\in\mathcal{K}_j}\mathbb{E}\left[\bar{R}_{c_j,k}\right]$. On the other hand, the EPR, $R_{e,p}=\sum_{l=1}^K \mathbb{E}\left[\bar{R}_l\right]$. The ESR of the proposed RS-CF system becomes}
\begin{equation}
     S_e=\sum_{j=1}^{N_c}\min_{k\in \mathcal{K}_j}\mathbb{E}\left[\bar{R}_{c_j,k}\right]+\sum_{l=1}^K \mathbb{E}\left[\bar{R}_l\right].\label{system ergodic sum rate}
 \end{equation}

\begin{remark}
The RS-CF is affected by the interference of multiple common symbols. This is expressed in \eqref{instantaneous SINR common rate} and \eqref{instantaneous SINR private rate} by the term $\sum\limits_{j=1,~j\neq i}^{N_c}a_{c_j}^2\lvert\boldsymbol{g}_k^{\normalfont{\text{T}}}\mathbf{p}_{c_j}\rvert^2$. However, when the clusters are properly designed they yield reduced levels of interference.
\end{remark}

 \begin{remark}
Under perfect CSIT, the instantaneous sum-rate is achievable. In such cases, $\mathbf{\tilde{G}}=\mathbf{0}$. The SINRs in \eqref{instantaneous SINR common rate} and \eqref{instantaneous SINR private rate} become, respectively,\par\noindent\small
\begin{align}
\gamma_{c_i,k}&=\frac{a_{c_i}^2\lvert \boldsymbol{g}_k^{\normalfont{\textrm{T}}}\mathbf{p}_{c_i}\rvert^2}{\sum\limits_{\substack{j=1\\j\neq i}}^{N_c}a_{c_j}^2\lvert\boldsymbol{g}_k^{\normalfont{\text{T}}}\mathbf{p}_{c_j}\rvert^2+\sum\limits_{l=1}^K a_l^2\lvert \boldsymbol{g}_k^{\normalfont{\textrm{T}}}\mathbf{p}_l\rvert^2+\sigma_w^2}\label{instantaneous SINR common rate perfect},\end{align}\normalsize
and\par\noindent\small
\begin{align}
\gamma_k&=\frac{a_k^2\lvert\boldsymbol{{g}}_k^{\normalfont{\textrm{T}}}\mathbf{p}_k\rvert^2}{\sum\limits_{\substack{j=1\\j\neq i}}^{N_c}a_{c_j}^2\lvert\boldsymbol{g}_k^{\normalfont{\text{T}}}\mathbf{p}_{c_j}\rvert^2+\sum\limits_{\substack{l=1\\l\neq k}}^K a_l^2\lvert\boldsymbol{g}_k^{\normalfont{\textrm{T}}}\mathbf{p}_l\rvert^2+\sigma_w^2}.\label{instantaneous SINR private rate perfect}
 \end{align}\normalsize
 \end{remark}
\begin{remark}
Under perfect CSIT, the ESR is obtained by substituting \eqref{instantaneous SINR common rate perfect} and \eqref{instantaneous SINR private rate perfect} in \eqref{instantaneous sum-rate} and computing $S_{e,p}=\mathbb{E}\left[S_i\right]$.
 \vspace{-1em}
 \end{remark}

\vspace{-1em}
\section{AP Selection and Cluster Precoder Design }
\label{sec:ap}

We present the AP selection strategy to determine the channel coefficients that need to be estimated, the method to form the clusters, and the design of the common and private precoders for the proposed RS-based CF system. Both precoders are chosen to be linear because of their simplicity and widespread use in RS and CF networks.
 \vspace{-1.5em}

\subsection{Jointly AP and User clustering}
Consider a cluster of APs $\mathcal{A}_i$ that serves a cluster of users $\mathcal{K}_i$. Assume the system forms $N_c$ clusters. The design of the clusters $\mathcal{A}_i$ and $\mathcal{K}_i$ is fundamental to achieving the best performance possible of the system. The performance metric is the sum-rate. Hence, the design of the cluster reduces to solving the following optimization problem:
\begin{equation}
    \max_{\mathcal{A}_i,\mathcal{K}_i,N_c} S_a\left(\mathcal{A}_i,\mathcal{K}_i, N_c\right).\label{Joint AP UE desing}
\end{equation}
Solving \eqref{Joint AP UE desing} is not trivial and would require obtaining the small-scale fading coefficients to guarantee that the optimum is achieved. Such a design is not appropriate for practical systems and is going to be addressed in future works. Instead, we propose a simple clustering algorithm that has low computational complexity and is suited for practical systems.

\subsection{AP Selection Strategy}
Consider the set $\mathcal{B}$ that contains the channel coefficients that should be estimated. The channel coefficients that belong to $\mathcal{B}$ are selected based on the large-scale fading coefficient. Define the mean as $\mu_{\zeta}=\frac{1}{MK}\sum\limits_{m=1}^{M}\sum\limits_{k=1}^{K}\zeta_{m,k}$. Then, each large-scale fading coefficient is normalized by applying the rule
\begin{equation}
    \zeta'_{m,k}=\zeta_{m,k}-\mu_\zeta.\label{Proposed Threshold}
\end{equation}
The parameter $\zeta'_{m,k}$ defines a threshold to select the coefficients in $\mathcal{B}$, establishing which symbols may be transmitted by each AP. If $\zeta'_{m,k}>0$, then the channel coefficient associated with $\zeta_{m,k}$ i.e., $g_{m,k}$, belongs to $\mathcal{B}$. Otherwise, the coefficient $g_{m,k}$ is ignored and the equivalent channel coefficient $\overline{\text{g}}_{m,k}$ 
is set to zero. The rationale of this strategy is that the largest $\zeta'_{m,k}$ coefficients are contained in the interval $\left(0,+\infty\right]$. 
Other coefficients encapsulate bad channel conditions for transmission. Therefore, their contribution to the received signal is small and can be discarded.

Following \cite{Palhares2020}, one may also choose a simplified approach in that the users are served by a fixed number $n_s$ of APs. To this end, the $n_s$ largest coefficients in $\boldsymbol{\zeta}_k=\left[\zeta_{1,k},\zeta_{2,k},\cdots,\zeta_{N,k}\right]^{\text{T}}$ are found such that no user is poorly served. However, depending on the selected $n_s$, a few small $g_{m,k}$ values may also get included in $\mathcal{B}$, thereby wasting resources. Alternatively, some large $g_{m,k}$ values may be discarded leading to severe degradation in the quality of the received signal.

\subsection{Design of Clusters}
Let us define the matrix $\mathbf{J}=\left[\mathbf{j}_1,\mathbf{j}_2,\cdots,\mathbf{j}_K\right] \in \mathbb{R}^{M\times K}$, which specifies the antennas that serve each user. In particular, the $(m,k)$-th element of $\mathbf{J}$ is
\begin{equation}
j_{m,k}=\begin{cases}
1, &\zeta'_{m,k}>0\\
0, &\text{otherwise.}
\end{cases}
\end{equation}
In other words, the $m$-th AP has a good channel quality and is a candidate to serve the $k$-th user if $j_{m,k}=1$. It follows that the vector $\mathbf{j}_k^{\textrm{T}}$ specifies those APs that serve the $k$-th user. As explained in the sequel, we assign the $k$-th user to a cluster based on the vector $\mathbf{j}_k^{\text{T}}$.

Consider the first cluster $\mathcal{K}_1$ that includes user $1$ and define the test vector of cluster $\mathcal{K}_1$ as $\bar{\mathbf{j}}_1^{\text{T}}=\mathbf{j}_1^{\text{T}}$. Then, we evaluate if user $2$ should be included in $\mathcal{K}_1$. To this end, we employ the test vector previously defined and compute the number $N_{a_{2,1}}$ of antennas that user 2 and the users in $\mathcal{K}_1$ share, i.e., $N_{a_{2,1}}=\mathbf{j}_2^{\text{T}}\bar{\mathbf{j}}_1$. If the parameter $N_{a_{2,1}}$ exceeds a predefined threshold $N_a$, we include user $2$ in $\mathcal{K}_1$ and update the test vector as $\bar{\mathbf{j}}_1^{\text{T}}\gets\bar{\mathbf{j}}_1^{\text{T}}\odot\mathbf{j}_2^{\text{T}}$. Otherwise, if $N_{a_{2,1}}<N_a$ we assign user $2$ to a new cluster, say cluster $\mathcal{K}_2$, and set $\mathbf{j}_2$ as its test vector. 

 The remaining users are assigned to a cluster following a similar procedure. For instance, while assigning any $k$-th user to a cluster, the system computes $N_{a_{k,i}}$, with $i=1,2,\cdots,N_c$. If $N_{a_{k,i}}> N_a$, then user $k$ belongs to $\mathcal{K}_i$ and the test vector of the $i$-th cluster is updated as $\bar{\mathbf{j}}_i^{\text{T}}\gets\bar{\mathbf{j}}_i^{\text{T}}\odot\mathbf{j}_k^{\text{T}}$. Otherwise, a new cluster is created, i.e.,  $N_c\gets N_c+1$ and the $k$-th user is allocated to the cluster $\mathcal{K}_{N_c}$ with the test vector  $\bar{\mathbf{j}}^{\text{T}}_{N_c}\gets\mathbf{j}^{\text{T}}_k$. Algorithm \ref{alg:cluster_design} summarizes our cluster design procedure. For every user $k$, the inner loop of Algorithm \ref{alg:cluster_design} evaluates if this user should be included in an existing cluster and the outer loop tests if the same user was assigned successfully to any cluster. If user $k$ was not assigned to any cluster, a new cluster is created. Finally, vector $\bar{\mathbf{j}}_i^{\textnormal{T}}$ defines the APs serving the \textit{i}-th cluster.

 With the users in $\mathcal{K}_k$ defined, we take the associated rows in $\overline{\mathbf{G}}$ to form the reduced matrix $\overline{\mathbf{G}}^{\text{T}}_k=\mathbf{U}_k\overline{\mathbf{G}}^{\text{T}}$ to compute the precoder with reduced dimensions. Here, the user selection matrix $\mathbf{U}_{k}\in\mathbb{R}^{\lvert\mathcal{K}_k\rvert \times K}$, whose first row 
has all zeros
except the $i$-th entry, where $i=\min\limits_{q\in\mathcal{K}_k}q$, i.e.,
\begin{equation}
\mathbf{u}_{1,k}=[\underbrace{0,\cdots,}_{i-1~~\textrm{terms}}1,\cdots,0] \in\mathbb{R}^{K}.
\end{equation}
The second row $\mathbf{u}_{2,k}$ contains all zeros 
except at the $l$-th entry, where $l$ is the second lowest index in $\mathcal{K}_k$. The following rows are obtained similarly.

{
\begin{algorithm}[H]
    \caption{Cluster design for RS-CF architecture }
    \label{alg:cluster_design}
    \begin{algorithmic}[1]
    \Statex \textbf{Input:}  $\mathbf{J}$, $N_a$
    \Statex \textbf{Output:} $\left\{\mathcal{K}_{i}\right\}_{i=1}^{N_c}$ 
\State $N_{c}\gets 1, \mathcal{K}_{i}\gets \left\{1\right\},\bar{\mathbf{j}}_1^{\text{T}}=\mathbf{j}_1^{\text{T}}$
\For{$k=2:K$}
\For{$i=1:N_c$}
\State $N_{a_{k,i}}=\mathbf{j}_k^{\text{T}}\bar{\mathbf{j}}_i$
\If{$N_{a_{k,i}}\geq N_a$}
\State $\bar{\mathbf{j}}_i^{\text{T}}\gets\bar{\mathbf{j}}_i^{\text{T}}\odot\mathbf{j}_k^{\text{T}}$
\State $\mathcal{K}_i\gets\mathcal{K}_i\cup\left\{k\right\}$
\State \textbf{break}
\EndIf
\State \textbf{end if}
\EndFor
\State \textbf{end for}
\If{$N_{a_{k,i}}<N_a$}
\State $N_c\gets N_c +1$
\State $\mathcal{K}_{N_c}\gets \left\{k\right\}$
\State $\bar{\mathbf{j}}^{\text{T}}_{N_c}=\mathbf{j}_k^{\text{T}}$
\EndIf
\State \textbf{end if}
\EndFor
\State \textbf{end for}
\Statex \Return $\left\{\mathcal{K}_{i}\right\}_{i=1}^{N_c}$
\end{algorithmic}
\end{algorithm}
}

{A fixed number of clusters $N_c$ are obtained by finding $N_c$ users that share the smallest number of antennas. Then, we accordingly update the test vector of each cluster and employ Algorithm 1 to allocate the remaining users.}

\vspace{-1.5em}
\subsection{Common Precoder}
Assume that the antenna selection policy produces a total of $N_c$ disjoint clusters, which results in $N_c$ common streams. Then, the common precoder $\mathbf{P}_c=\left[\mathbf{p}_{c_1},\mathbf{p}_{c_2},\cdots,\mathbf{p}_{c_{N_c}}\right]$ maps the common symbols to their respective transmit antennas. Note that the symbol $s_{c_i}$ contains only information from the users in $\mathcal{K}_{i}$. Therefore, only the users in $\mathcal{K}_{i}$ must decode it. Users in $\mathcal{K}_j, \forall j\neq i$ treat $s_{c_i}$ as residual interference. Note that the channel coefficients $\overline{g}_{m,k}$, 
with $m \notin\mathcal{A}_i $, were tagged as inappropriate for transmission of the common message. Therefore, the residual interference caused by $s_{c_k}$ at clusters $\mathcal{K}_i, \forall i\neq k$ is mitigated because of the unfavorable propagation conditions.

Let us consider the reduced channel matrix $\overline{\mathbf{G}}_k$ defined in the last section since only the user in $\mathcal{K}_{k}$ is important for the transmission of $s_{c_k}$. Once $\overline{\mathbf{G}}_k$ is defined, we compute the common precoder by applying a singular value decomposition over the reduced channel, i.e., $\overline{\mathbf{G}}_k=\overline{\mathbf{U}}_k\overline{\mathbf{\Psi}}_k\overline{\mathbf{V}}^{H}_k$. The matrices $\overline{\mathbf{U}}_k$ and $\overline{\mathbf{V}}_k$ are complex unitary matrices and $\overline{\mathbf{\Psi}}_k$ is a diagonal matrix containing the singular values of $\overline{\mathbf{G}}_k$. Then, we set the common precoder $\mathbf{p}_{c_k}=\overline{\mathbf{v}}_1^{\left(k\right)},\; \forall k=1,2,\cdots,N_c$. Finally, the common precoder is given by $\mathbf{P}_c=\left[\overline{\mathbf{v}}_1^{\left(1\right)},\overline{\mathbf{v}}_1^{\left(2\right)},\cdots,\overline{\mathbf{v}}_1^{\left(N_c\right)}\right].$

\subsection{Private Precoders}
The RS employs separate precoders for private streams. We consider the following private precoders for the proposed RS-CF architecture. Unlike standard RS private precoders, we take the AP selection strategy and cluster formation into account.
\subsubsection{MMSE Precoder}
We use the equivalent channel estimate $\overline{\mathbf{G}}$ to compute the MMSE precoder and reduce the signaling load. This sparse MMSE (MMSE-SP) precoder is given by
\begin{equation}
    \mathbf{P}^{\left(\text{MMSE-SP}\right)}=\beta^{\left(\text{MMSE}\right)}\underbrace{\overline{\mathbf{G}}^*\left(\overline{\mathbf{G}}^{\text{T}}\overline{\mathbf{G}}^*+\frac{K\sigma_w^2}{P_t}\mathbf{I}_{K}\right)^{-1}}_{\mathbf{F}^{\left(\text{MMSE}\right)}} ~~ \in{\mathbb{C}}^{M \times K},\label{MMSE precoder full network}
\end{equation}
where $\beta^{\left(\text{MMSE}\right)}=\sqrt{P_{t}/\text{tr}\left\{\mathbf{F}^{\left(\text{MMSE}\right)}\mathbf{F}^{\left(\text{MMSE}\right)^H}\right\}}$.


Although \eqref{MMSE precoder full network} saves system resources, the matrix inversion step is computationally demanding. To address this, we propose the MMSE with reduced dimensions (MMSE-RD) 
The MMSE-RD precoder can be obtained by solving the following optimization problem:
\begin{gather}
    \left\{\mathbf{P}_k^{\left(\text{MMSE-RD}\right)},\beta^{\left(\text{MMSE-RD}\right)}\right\}=\min_{\mathbf{P}_k,\beta}\mathbb{E}\left[\lVert\mathbf{s}_k-\beta\mathbf{y}_k\rVert^2\right]\nonumber\\
    \textrm{subject to }\mathbb{E}\left[\lvert\mathbf{x}_k\rvert^2\right]=P_{t}/K,\label{MMSE optimization problem}
\end{gather}
where $\mathbf{s}_k=\mathbf{U}_k\mathbf{s}\in\mathbb{C}^{\lvert\mathcal{K}_k\rvert}$, $\mathbf{s}=\left[s_1,\cdots,s_K\right]^{\text{T}}\in \mathbb{C}^{K}$ and $\mathbf{y}_k=\overline{\mathbf{G}}_k\mathbf{x}_k+\mathbf{U}_k\mathbf{w}\in\mathbb{C}^{\lvert\mathcal{K}_k\rvert}$ contain, respectively, the symbols and the received signal of the users in $\mathcal{K}_k$; and $\mathbf{x}_k=\mathbf{P}_k\mathbf{s}_k\in\mathbb{C}^{M}$ is the transmitted signal of the users in $\mathcal{K}_k$. Following a similar procedure as the one described in \cite{Joham2005}, we find the solution to the optimization problem in \eqref{MMSE optimization problem} as 
\begin{equation}
    \mathbf{P}_k^{\left(\text{MMSE-RD}\right)}=\beta^{\left(\text{MMSE-RD}\right)}\bar{\mathbf{P}}_k \;\;\in \mathbb{C}^{M\times\lvert\mathcal{K}_k\rvert}, \label{MMSE subset}
\end{equation}
where
\begin{equation}
   \beta^{\left(\text{MMSE-RD}\right)}=\sqrt{\frac{P_{t}}{K\text{tr}\left\{\bar{\mathbf{P}}_k\bar{\mathbf{P}}^H_k\right\}}},
\end{equation}
\begin{equation}
  \bar{\mathbf{P}}_k=\overline{\mathbf{G}}^*_k\left(\overline{\mathbf{G}}^{\text{T}}_{k}\overline{\mathbf{G}}_k^{\text{*}}+\frac{K\lvert\mathcal{K}_k\rvert\sigma_w^2}{P_{t}}\mathbf{I}_{\lvert\mathcal{K}_k\rvert}\right)^{-1},
\end{equation}
where $\sigma_w^2$ is the variance of the AWGN. 

We note that the set $\mathcal{K}_k$ is associated with the decoding of the symbol $s_k$. Since the matrix dimensions have been reduced, we need to employ index mapping to obtain the appropriate precoder for $s_k$. Given that the $k$-th stream will be decoded, we find the row $\mathbf{u}_{q,k}$ that contains a one in its $k$-th entry. It follows that the $q$-th column of
$\mathbf{P}_k^{\left(\text{MMSE-RD}\right)}$ maps the symbol $s_k$ and  is, therefore, employed in $\mathbf{P}^{\left(\text{RU-MMSE-RD}\right)}= [\mathbf{p}_1^{\left(\text{RU-MMSE-RD}\right)} \ldots \mathbf{p}_k^{\left(\text{RU-MMSE-RD}\right)} \ldots \mathbf{p}_{K}^{\left(\text{RU-MMSE-RD}\right)} ] \in \mathbb{C}^{M\times K}$ i.e.,
\begin{equation}
    \mathbf{p}_k^{\left(\text{RU-MMSE-RD}\right)}=\left[\mathbf{P}_{k}^{\left(\text{MMSE-RD}\right)}\right]_q.
\end{equation}
Note that only the users in $\mathcal{K}_k$ are considered when computing the column vector $\mathbf{p}^{\left(\text{RU-MMSE-RD}\right)}_k$ and a small residual interference remains.

\subsubsection{ZF precoder design}
A particular instance of the MMSE precoder is the ZF precoder, which is designed to remove the MUI completely. The \textit{sp}arse ZF (ZF-SP) similarly reduces the signaling load by employing a sparse matrix precoder and is computed using the pseudoinverse \cite{Joham2005} as
\begin{equation}
    \mathbf{P}^{\left(\text{ZF-SP}\right)}=\beta^{\left(\text{ZF}\right)}\underbrace{\overline{\mathbf{G}}^{*}\left(\overline{\mathbf{G}}^{\text{T}}\overline{\mathbf{G}}^{*}\right)^{-1}}_{\mathbf{F}^{\left(\text{ZF}\right)}} ~~ \in{\mathbb{C}}^{M \times K},\label{ZF network wide}
\end{equation}
where $\beta^{\left(\text{ZF}\right)}=\sqrt{P_{t}/\text{tr}\left\{\mathbf{F}^{\left(\text{ZF}\right)}\mathbf{F}^{\left(\text{ZF}\right)^H}\right\}}$.


The dimension of the matrix that requires inversion may also be reduced by considering only those users that belong to $\mathcal{K}_k$. This yields the reduced dimension ZF precoder as 
\begin{equation}
    \mathbf{P}^{\left(\text{ZF-RD}\right)}_k=\overline{\mathbf{G}}_k^*\left(\overline{\mathbf{G}}^{\text{T}}_k\overline{\mathbf{G}}_k^*\right)^{-1}\;\; \in \mathbb{C}^{M\times\lvert\mathcal{K}_k\rvert}.
\end{equation}
Then, we find the row $\mathbf{u}_{q,k}$ that has one as its $k$-th entry. It follows that the $k$-th column of the cluster-based restricted-users ZF precoder $\mathbf{P}^{\left(\text{RU-ZF-RD}\right)} \in\mathbb{C}^{M\times K} $ is equal to the $q$-th column of $\mathbf{P}_k^{\left(\text{ZF-RD}\right)}$, i.e.,
\begin{equation}
    \mathbf{p}^{\left(\text{RU-ZF-RD}\right)}_k=\left[\mathbf{P}_{k}^{\left(\text{ZF-RD}\right)}\right]_q.\label{ZF subset precoder}
\end{equation}

\subsubsection{MF precoder design}
We employ this precoder to send the private symbols because of its reduced computational complexity. This precoder is readily obtained by the Hermitian of the \textit{sp}arse channel estimate \cite{Joham2005}, i.e.,
\begin{equation}
\mathbf{P}^{\left(\text{MF-SP}\right)}=\overline{\mathbf{G}}^*. \label{MF-SP precoder}
\end{equation}
\section{Joint RS-CF Resource Allocation}
\label{sec:alloc}
The overall performance of the proposed RS-CF architecture depends on appropriate resource allocation. In this context, power allocation is a critical step to harness the RS-CF benefits. Specifically, the gain obtained in the sum-rate performance, which is the figure-of-merit  adopted in this work to evaluate the proposed RS-CF architecture, depends on the power allocation. Unlike other works, the proposed RS-CF architecture sends one common message per cluster. This makes the problem challenging because multiple common streams should be taken into account for power allocation. 

The proposed RS-CF architecture must fulfill the transmit power constraint, i.e., $\mathbb{E}\left[\lVert\mathbf{x}^{\left(\text{RS}\right)}\rVert^2\right]\leq P_t$. Note that part of the available power must be reserved to transmit the common symbols. In particular, we have $a_{c_i}^2=\delta_i P_t$,
where $\delta_i\in \left[0,1\right],~~\forall i,$ represents the fraction of power allocated to the common stream $s_{c_i}$. Furthermore, $\delta=\sum\limits_{i=1}^{N_c}\delta_i <1$. Then, the power available for transmitting the private streams is $\left(1-\sum_{i=1}^{N_c}\delta_i\right)P_t$.

For the proposed cluster-based RS-CF architecture, an appropriate power allocation amounts to finding an appropriate value of the vector $\boldsymbol{\delta}=\left[\delta_1,\delta_2,\cdots,\delta_{N_c}\right]^{\text{T}}\in\mathbb{R}^{N_c}$, which contains the fraction of available power that should be assigned to each common stream of each cluster. If $\mathbf{\delta}$ is properly set, the performance of RS-CF should be at least as good as the performance obtained by CF without RS because the system can set $\boldsymbol{\delta}=\mathbf{0}$, which implies that the splitting procedure is avoided resulting in a 
conventional CF system. 

Denote the optimal value of $\boldsymbol{\delta}$ given the $i$-th channel estimate by $\boldsymbol{\delta}^{\left(o\right)}_i$. To obtain the best power allocation for the common streams, we need to find the sequence of $\boldsymbol{\delta}^{\left(o\right)}_i$ that maximizes the ESR. Thus, considering $L$ channel realizations, we have $\boldsymbol{\Delta}^{\left(o\right)}=\left[\boldsymbol{\delta}^{\left(o\right)}_1,\boldsymbol{\delta}^{\left(o\right)}_2,\cdots,\boldsymbol{\delta}^{\left(o\right)}_L\right] \in \mathbb{C}^{N_c \times L}$, which is the solution of the following optimization problem:
\begin{equation}
    \boldsymbol{\Delta}^{\left(o\right)}= \max_{\boldsymbol{\Delta}}S_e\left(\boldsymbol{\Delta}\right),\label{ESR optimization problem}
\end{equation}

The optimization problem in \eqref{ESR optimization problem} is non-convex and depends on the sequence of channel realization. 
Therefore, we relax the problem and consider a sequence of maximization problems considering a single channel realizations, i.e., maximizing the ASR as
\begin{equation}
    \boldsymbol{\delta}^{\left(o\right)}_i= \max_{\boldsymbol{\delta}_i} S_a\left(\boldsymbol{\delta}_i\right).\label{ASR optimization problem}
\end{equation}
Although the ESR obtained with \eqref{ESR optimization problem} can be higher than \eqref{ASR optimization problem}, the latter has a greater impact on the tractability of the problem because the power allocation is performed independently for each channel realization.
The solution of \eqref{ASR optimization problem} can be found by employing an exhaustive search to find a suitable value $\boldsymbol{\delta}^{\left(o\right)}_i$ given that a predefined fixed power allocation is used across the private symbols. However, as the number of clusters grows, the computational cost may exponentially grow. To simplify this, we enforce $a_{c_i}=a_{c_j}~~\forall i,j$, i.e., we allocate the same power to the common streams. Then, the problem is reduced to finding the optimal $\delta=N_c \delta_i$ that solves
\small{\begin{align}
    \underset{\delta}{\textrm{maximize}} & \left(\sum_{j=1}^{N_c}\min_{k}\bar{R}_{c_j,k}\left(\delta\right)+\sum_{k=1}^{K}\bar{R}_k\left(\delta\right)\right)\nonumber\\
    \text{subject to } &  \mathbb{E}\left[\lVert\mathbf{x}^{\left(\text{RS}\right)}\rVert^2\right]\leq P_t,\nonumber\\
    & a_{c_i}=a_{c_j}~~\forall~{i,j}.\label{Power allocation problem}
\end{align}}
\normalsize
 Algorithm~\ref{alg:power} summarizes the steps to solve \eqref{Power allocation problem}. Define the step size $\mu$ to search for the optimal value of $\delta$. First, we set $\delta=0$ and the optimal power coefficients of the common streams equal to $\mathbf{a}_c^{\left(o\right)}=\mathbf{0}$. This corresponds to a conventional CF-MIMO system, where power is not allocated to the common stream. Then, the average sum-rate is computed and stored in $S_a$.  Next, we increase $\delta$ by $\mu$ and compute the available common power with $P_{t,c}=\delta P_t$. The result is equally distributed among the common streams and the average sum-rate is computed again. If the new value of $S_a$ is greater than the previous, we set $\mathbf{a}_c^{\left(o\right)}=\frac{P_{t,c}}{N_c}\mathbf{1}_{N_c}$. Then, we update the value of $\delta$ by adding $\mu$ and repeat the procedure until $\delta=1$ is reached. Other approaches such as convex optimization \cite{JoudehClerckx2016} and monotonic optimization \cite{Tuy2000} may also be employed for power allocation in the proposed RS-CF.

\begin{algorithm}[H]
    \caption{Power allocation for common streams }
    \label{alg:power}
    \begin{algorithmic}[1]
    \Statex \textbf{Input:}  $\mathbf{a}_p$, $\mu$, $\hat{\mathbf{H}}$
    \Statex \textbf{Output:} $\mathbf{a}_{c}^{\left(o\right)}$ 
\State $S_{a}\gets 0, \mathbf{a}_{c}^{\left(o\right)}\gets 0$
\For{$\delta=0:\mu:1$}
\State $P_{t,c}\gets\delta P_t$
\State $\mathbf{a}_{c}\gets\frac{P_{t,c}}{N_c}\mathbf{1}_{N_c}$
$\triangleright$ \textrm{Common power equally distributed among clusters} 
\State $S_{a,1}\gets\min_{k}\bar{R}_{c,k}\left(\mathbf{a}_{c}\right)+\sum_{k=1}^{K}\bar{R}_k\left(\mathbf{a}_{c}\right)$
\If{$S_{a}<S_{a,1}$}
\State $S_{a}\gets S_{a,1}$
\State $\mathbf{a}_{c}^{\left(o\right)}\gets \mathbf{a}_c$
\EndIf
\EndFor
\Statex \Return $\mathbf{a}_{c}^{\left(o\right)}$

\end{algorithmic}
\end{algorithm}
 \vspace{-2em}
\section{Performance Analyses} 
\label{sec:perf}
We present a sum-rate analysis of the proposed RS-CF scheme and cluster-based linear precoders, including a study of their computational complexity and signaling load.
\vspace{-1em}

 \subsection{Sum-rate of MF precoder}
To obtain the ESR for the MF precoder, we derive a closed-form expression for the SINR in the following Proposition~\ref{prop:sinr_MF}.
\begin{proposition}
\label{prop:sinr_MF}
When decoding the common stream, the SINR of the proposed MF-SP for RS-CF systems is
\par\noindent\small
\begin{equation}
     \gamma_{c_i,k}=\frac{a_{c_i}^2\psi_{1,1}^{\left(i\right)^2}\lvert u_{k,1}^{\left(i\right)}\rvert^2}{d_{c_i,k}^{\left(\text{v}\right)}+\sum\limits_{\substack{j=1\\j\neq i}}^{N_c} a_{c_j}^2\lvert\boldsymbol{g}_k^{\normalfont{\textrm{T}}}\mathbf{v}_{1}^{\left(j\right)}\rvert^2+\sum\limits_{l=1}^K a_l^2\lvert\boldsymbol{g}^{\normalfont{\textrm{T}}}_k\hat{\boldsymbol{g}}_l^*\rvert^2+\sigma_w^2/\varepsilon^2},\label{SINR when decoding common message MF approach}
 \end{equation}\normalsize
  where $d_{c_i,k}^{\left(\text{v}\right)}=a_{c_i}^2\left(2\psi_{1,1}^{\left(i\right)}\Re\left\{u_{k,1}^{\left(i\right)^*}\left(\tilde{\boldsymbol{g}}_k^{\normalfont{\textrm{T}}}\mathbf{v}_1^{\left(i\right)}\right)\right\}+\lvert\tilde{\boldsymbol{g}}^{\normalfont{\textrm{T}}}_k\mathbf{v}_1^{\left(i\right)}\rvert^2\right)$.

  When decoding the private message after SIC, the SINR is
  \par\noindent\small
    \begin{equation}
     \gamma_k=\frac{a_k^2\lVert\hat{\boldsymbol{g}}_k\rVert^4}{d_{k}^{\left(\text{MF}\right)}+\sum\limits_{\substack{j=1\\j\neq i}}^{N_c} a_{c_j}^2\lvert\boldsymbol{g}^{\normalfont{\textrm{T}}}_k\mathbf{v}_{j}^{\left(i\right)}\rvert^2+\sum\limits_{\substack{l=1\\l\neq k}}^K a_l^2\lvert\boldsymbol{g}^{\normalfont{\textrm{T}}}_k\tilde{\boldsymbol{g}}_l^*\rvert^2+\sigma_w^2/\varepsilon^2},\label{SINR when decoding private message MF approach}
 \end{equation}\normalsize
   where $d_{k}^{\left(\text{MF}\right)}=a_k^2\left( 2\lVert\hat{\boldsymbol{g}}_k\rVert^2\Re\left\{\tilde{\boldsymbol{g}}_k^{\normalfont{\textrm{T}}}\hat{\boldsymbol{g}}^*_k\right\}+\lvert\tilde{\boldsymbol{g}}_k^{\text{T}}\hat{\boldsymbol{g}}^*_k\rvert^2\right)$.
 \end{proposition}
 \begin{IEEEproof}
 Let us consider the MF defined in \eqref{MF-SP precoder} as the private precoder. Then, we can compute the average power of the received signal at user $k$ in $\mathcal{K}_i$ as \par\noindent\small
 \begin{align}
     \mathbb{E}\left[\lvert y_k\rvert^2\right]=&\varepsilon^2\left(a_{c_i}^2\underbrace{\lvert\left(\hat{\boldsymbol{g}}_k^{\text{T}}+\tilde{\boldsymbol{g}}^{\text{T}}_k\right)\mathbf{v}_{1}^{\left(i\right)}\rvert^2}_{T_1}+\sum\limits_{\substack{j=1\\j\neq i}}^{N_c} a_{c_j}^2\lvert\left(\hat{\boldsymbol{g}}_k^{\text{T}}+\tilde{\boldsymbol{g}}^{\text{T}}_k\right)\mathbf{v}_{1}^{\left(j\right)}\rvert^2+\sum_{l=1}^K a_l^2\underbrace{\lvert\left(\hat{\boldsymbol{g}}_k^{\text{T}}+\tilde{\boldsymbol{g}}^{\text{T}}_k\right)\hat{\boldsymbol{g}}_l^*\rvert^2}_{T_2}\right)+\sigma_w^2.\label{average receive power MF plus SVD}
 \end{align}
 \normalsize
 Expanding $T_1$ gives\par\noindent\small
 \begin{align}
     T_1=&\left(\hat{\boldsymbol{g}}^{\text{T}}_k\mathbf{v}_{1}^{\left(i\right)}+\tilde{\boldsymbol{g}}^{\text{T}}_k\mathbf{v}_{1}^{\left(i\right)}\right)^*\left(\hat{\boldsymbol{g}}^{\text{T}}_k\mathbf{v}_{1}^{\left(i\right)}+\tilde{\boldsymbol{g}}^{\text{T}}_k\mathbf{v}_{1}^{\left(i\right)}\right)\nonumber\\
     =&\lvert\hat{\boldsymbol{g}}^{\text{T}}_k\mathbf{v}_{1}^{\left(i\right)}\rvert^2+2\Re\left\{\left(\hat{\boldsymbol{g}}^{\text{T}}_k\mathbf{v}_{1}^{\left(i\right)}\right)^*\left(\tilde{\boldsymbol{g}}^{\text{T}}_k\mathbf{v}_{1}^{\left(i\right)}\right)\right\}+\lvert\tilde{\boldsymbol{g}}^{\text{T}}_k\mathbf{v}_{1}^{\left(i\right)}\rvert^2.\label{T1 evaluation}
 \end{align}\normalsize
Note that \par\noindent\small
 \begin{align}
     \hat{\boldsymbol{g}}^{\text{T}}_k\mathbf{v}_{1}^{\left(i\right)}= \mathbf{u}_{k,*}^{\left(i\right)}\boldsymbol{\Psi}_i\mathbf{V}^H_i\mathbf{v}_{1}^{\left(i\right)}
     =\mathbf{u}_{k,*}^{\left(i\right)}\boldsymbol{\psi}_{1}^{\left(i\right)}
     =u_{k,1}^{\left(i\right)}\psi_{1,1}^{\left(i\right)}.
 \end{align}\normalsize
Then, \eqref{T1 evaluation} becomes\par\noindent\small
\begin{equation}
    T_1=\psi_{1,1}^{\left(i\right)^2}\lvert u_{k,1}^{\left(i\right)}\rvert^2+2\psi_{1,1}^{\left(i\right)}\Re\left\{u_{k,1}^{\left(i\right)^*}\left(\tilde{\boldsymbol{g}}_k^{\text{T}}\mathbf{v}_{1}^{\left(i\right)}\right)\right\}+\lvert\tilde{\boldsymbol{g}}^{\text{T}}_k\mathbf{v}_{1}^{\left(i\right)}\rvert^2.\label{T1 reduced}
\end{equation}\normalsize

Expanding $T_2$
yields\par\noindent\small
\begin{align}
    T_2=\lvert\hat{\boldsymbol{g}}^{\text{T}}_k\hat{\boldsymbol{g}}^*_l\rvert^2+2\Re\left\{\left(\hat{\boldsymbol{g}}^{\text{T}}_k\hat{\boldsymbol{g}}_l^*\right)^*\left(\tilde{\boldsymbol{g}}^{\text{T}}_k\hat{\boldsymbol{g}}_l^*\right)\right\}+\lvert\tilde{\boldsymbol{g}}^{\text{T}}_k\hat{\boldsymbol{g}}_l^*\rvert^2.\label{T2 expanded}
\end{align}\normalsize
When $k=l$, \eqref{T2 expanded} reduces to\par\noindent\small
\begin{equation}
    T_2=\lVert\hat{\boldsymbol{g}}_k\rVert^4+2\lVert\hat{\boldsymbol{g}}_k\rVert^2\Re\left\{\tilde{\boldsymbol{g}}_k^{\text{T}}\hat{\boldsymbol{g}}^*_k\right\}+\lvert\tilde{\boldsymbol{g}}_k^{\text{T}}\hat{\boldsymbol{g}}^*_k\rvert^2.\label{T2 reduced}
\end{equation}\normalsize

 By substituting \eqref{T1 reduced} and \eqref{T2 reduced} in \eqref{instantaneous SINR common rate} and \eqref{instantaneous SINR private rate}, we get \eqref{SINR when decoding common message MF approach} and \eqref{SINR when decoding private message MF approach}.

\end{IEEEproof}
The ESR of the proposed RS-CF system is obtained by substituting \eqref{SINR when decoding common message MF approach} and \eqref{SINR when decoding private message MF approach} into \eqref{instantaneous sum-rate}, \eqref{system average sum rate}, and \eqref{system ergodic sum rate}. 

 \begin{remark}
 Under perfect CSIT, the SINRs simplify to\par\noindent\small
 \begin{equation}
     \gamma_{c_i,k}=\frac{a_{c_i}^2\psi_{1,1}^{\left(i\right)^2}\lvert u_{k,1}^{\left(i\right)}\rvert^2}{\sum\limits_{\substack{j=1\\j\neq i}}^{N_c} a_{c_j}^2\lvert\hat{\boldsymbol{g}}_k^{\normalfont{\textrm{T}}}\mathbf{v}_{1}^{\left(j\right)}\rvert^2+\sum\limits_{l=1}^K a_l^2\lvert\hat{\boldsymbol{g}}_k^{\normalfont{\textrm{T}}}\hat{\boldsymbol{g}}_l^*\rvert^2+\sigma_w^2},
 \end{equation}\normalsize
 and\par\noindent\small
 \begin{equation}
     \gamma_k=\frac{a_k^2\lVert\hat{\boldsymbol{g}}^{\normalfont{\textrm{T}}}_k\rVert^4}{\sum\limits_{\substack{j=1\\j\neq i}}^{N_c} a_{c_j}^2\lvert\hat{\boldsymbol{g}}_k^{\normalfont{\textrm{T}}}\mathbf{v}_{1}^{\left(i\right)}\rvert^2+\sum\limits_{\substack{i=1\\i\neq k}}^K a_i^2\lvert\hat{\boldsymbol{g}}_k^{\normalfont{\textrm{T}}}\hat{\boldsymbol{g}}_i^*\rvert^2+\sigma_w^2}.
 \end{equation}\normalsize
 \end{remark}

 \subsection{Sum-rate of ZF precoder}
Similar to the MF precoder, we have the following Proposition~\ref{prop:sinr_ZF Sparse} for the SINR of the ZF-SP.
 \begin{proposition}
 \label{prop:sinr_ZF Sparse}
  For the ZF-SP, the SINR when decoding the common message at the $k$-th user, which belongs to $\mathcal{K}_i$, is\par\noindent\small
 \begin{align}
     &\gamma_{c_i,k}=\frac{a_{c_i}^2\psi_{1,1}^{\left(i\right)^2}\lvert u_{k,1}^{\left(i\right)}\rvert^2}{d_{c_i,k}^{\left(v\right)}+\sum\limits_{\substack{j=1\\j\neq i}}^{N_c} a_{c_j}^2\lvert\boldsymbol{g}_k^{\normalfont{\text{T}}}\mathbf{v}_{1}^{\left(j\right)}\rvert^2+a_k^2+d_k^{\left(\normalfont{\textrm{ZF}}\right)} +\sum\limits_{l=1}^{K}a_l^2\lvert \tilde{\boldsymbol{g}}_k^{\text{T}}\normalfont{\hat{\mathbf{G}}}^{*}\boldsymbol{\lambda}_l\rvert^2+\sigma_w^2/\varepsilon^2},\label{SINR when decoding common message ZF approach}
 \end{align}\normalsize
where $d_k^{\left(\normalfont{\textrm{ZF}}\right)}=-2a_k^2\Re\left\{\tilde{\boldsymbol{g}}_k^{\normalfont{\text{T}}}\hat{\normalfont{\mathbf{G}}}^*\boldsymbol{\lambda}_k \right\}$.

When decoding the private message at $k$-th user, the SINR is \par\noindent\small
\begin{equation}
    \gamma_k=\frac{a_k^2}{d_k^{\normalfont{\left(\text{ZF}\right)}}+\sum\limits_{\substack{j=1\\j\neq i}}^{N_c} a_{c_j}^2\lvert\boldsymbol{g}_k^{\text{T}}\mathbf{v}_{1}^{\left(j\right)}\rvert^2+\sum\limits_{l=1}^{K}a_l^2\lvert\tilde{\boldsymbol{g}}_k^{\text{T}}\hat{\normalfont{\mathbf{G}}}^{*}\boldsymbol{\lambda}_l\rvert^2+\sigma_w^2/\varepsilon^2}.\label{SINR when decoding private message ZF approach}
\end{equation}\normalsize
 \end{proposition}
 \begin{IEEEproof}
  For the ZF precoder, we have $\hat{\boldsymbol{g}}_k^{\text{T}}\mathbf{p}_k=1$ and $\hat{\boldsymbol{g}}_k^{\text{T}}\mathbf{p}_i=0$ for all $i\neq k$. Then, the average power of the received signal is\par\noindent\small
 \begin{align}
     \mathbb{E}\left[\lvert y_k\rvert^2\right]=&\varepsilon^2\left(a_{c_i}^2 T_1+\sum\limits_{\substack{j=1\\j\neq i}}^{N_c} a_{c_j}^2\lvert\left(\hat{\boldsymbol{g}}_k^{\text{T}}+\tilde{\boldsymbol{g}}^{\text{T}}_k\right)\mathbf{v}_{1}^{\left(j\right)}\rvert^2+a_k^2-2a_k^2\Re\left\{\tilde{\boldsymbol{g}}_k^{\normalfont{\text{T}}}\hat{\normalfont{\mathbf{G}}}^*\boldsymbol{\lambda}_k \right\}+\sum_{l=1}^{K}a_l^2\lvert \tilde{\boldsymbol{g}}_k^{\text{T}}\hat{\mathbf{G}}^{*}\boldsymbol{\lambda}_l\rvert^2\right)+\sigma_w^2,\label{Average power of the received signal ZF approach}
 \end{align}\normalsize
 where $\boldsymbol{\Lambda}=\left(\hat{\mathbf{G}}^{\text{T}}\hat{\mathbf{G}}^*\right)^{-1}$ and the $l$-th column of $\boldsymbol{\Lambda}$ is the vector $\boldsymbol{\lambda}_l$. From \eqref{Average power of the received signal ZF approach}, we obtain \eqref{SINR when decoding common message ZF approach} and \eqref{SINR when decoding private message ZF approach}.
 \end{IEEEproof}

The ESR is computed by substituting \eqref{SINR when decoding common message ZF approach} and \eqref{SINR when decoding private message ZF approach} into \eqref{instantaneous common rate}, \eqref{instantaneous private rate}, and \eqref{system ergodic sum rate}.
The following proposition~\ref{prop:sinr_ZF} states the SINR for RU-ZF-RD.
 \begin{proposition}
 \label{prop:sinr_ZF}
  For the RU-ZF-RD, the SINR when decoding the common message at the $k$-th user in $\mathcal{K}_i$ is\par\noindent\small
 \begin{equation}
     \gamma_{c_i,k}=\frac{a_{c_i}^2\psi_{1,1}^{\left(i\right)^2}\lvert u_{k,1}^{\left(i\right)}\rvert^2}{d_{c_i,k}^{\left(v\right)}+\sum\limits_{\substack{j=1\\j\neq i}}^{N_c} a_{c_j}^2\lvert\boldsymbol{g}_k^{\normalfont{\text{T}}}\mathbf{v}_{1}^{\left(j\right)}\rvert^2+a_k^2+d_k^{\left(\normalfont{\text{RD}}\right)}+\sigma_w^2/\varepsilon^2},\label{SINR when decoding common message ZF-RD approach}
 \end{equation}\normalsize
where $d_k^{\left(\normalfont{\text{RD}}\right)}=d_k^{\left(\normalfont{\text{ZF}}\right)}+\sum\limits_{\substack{j=1\\j\neq i}}^{N_c}\sum\limits_{r\in \mathcal{K}_j} a_r^2\lvert\boldsymbol{g}^{\normalfont{\text{T}}}_k\hat{\mathbf{G}}^*_j\boldsymbol{\lambda}_{r\rightarrow q}^{\left(j\right)}\rvert^2 +\sum\limits_{\substack{l\in\mathcal{K}_i}}^{K}a_l^2\lvert \tilde{\boldsymbol{g}}_k^{\text{T}}\mathbf{\hat{G}}_l^{*}\boldsymbol{\lambda}^{\left(l\right)}_{l\rightarrow v}\rvert^2$ and $\boldsymbol{\Lambda}_{j}=\left(\hat{\mathbf{G}}^{\text{T}}_j\hat{\mathbf{G}}^{*}_j\right)^{-1}$, whose $l$-th column $\boldsymbol{\lambda}^{\left(j\right)}_{l\rightarrow q}$ is associated with the decoding of the $q$-th symbol.

The SINR when decoding the private message at the $k$-th user is \par\noindent\small
\begin{equation}
    \gamma_k=\frac{a_k^2}{\sum\limits_{\substack{j=1\\j\neq i}}^{N_c} a_{c_j}^2\lvert\boldsymbol{g}_k^{\normalfont{\text{T}}}\mathbf{v}_{1}^{\left(j\right)}\rvert^2+d_k^{\left(\normalfont{\text{RD}}\right)}+\sigma_w^2/\varepsilon^2}.\label{SINR when decoding private message ZF-RD approach}
\end{equation}\normalsize
 \end{proposition}
 \begin{IEEEproof}
 By squaring and taking the expected value of the received signal at the $k$-th user, we obtain
 \begin{align}
     \mathbb{E}\left[\lvert y_k\rvert^2\right]=&\varepsilon^2\left(a_{c_i}^2\psi_{1,1}^{\left(i\right)^2}\lvert u_{k,1}^{\left(i\right)}\rvert^2+d_{c_i,k}^{\left(v\right)}+\sum\limits_{\substack{j=1\\j\neq i}}^{N_c} a_{c_j}^2\lvert\boldsymbol{g}_k^{\normalfont{\text{T}}}\mathbf{v}_{1}^{\left(j\right)}\rvert^2\vphantom{\sum\limits_{\substack{j=1\\j\neq i}}^{N_c}}+a_k^2+d_k^{\left(\normalfont{\text{RD}}\right)}\right)+\sigma_w^2.\label{received signal power RU-ZF-RD}
 \end{align}
 From \eqref{received signal power RU-ZF-RD}, we obtain \eqref{SINR when decoding common message ZF-RD approach} and \eqref{SINR when decoding private message ZF-RD approach}.
 \end{IEEEproof}

\subsection{Sum-rate of MMSE precoder}
The following Proposition~\ref{prop:sinr_MMSE} states the SINR of the proposed RS-CF architecture with a linear MMSE precoder.
\begin{proposition}
 \label{prop:sinr_MMSE}
  For the MMSE precoder, the SINR when decoding the common message is\par\noindent\small
 \begin{equation}
     \gamma_{c_i,k}=\frac{a_{c_i}^2\psi_{1,1}^{\left(i\right)^2}\lvert u_{k,1}^{\left(i\right)}\rvert^2}{d_{c_i,k}^{\left(v\right)} +\sum\limits_{\substack{j=1\\j\neq i}}^{N_c} a_{c_j}^2\lvert\left(\hat{\boldsymbol{g}}_k^{\normalfont{\textrm{T}}}+\tilde{\boldsymbol{g}}^{\normalfont{\textrm{T}}}_k\right)\mathbf{v}_{1}^{\left(j\right)}\rvert^2+\sum\limits_{l=1}^{K}a_l^2\lvert \boldsymbol{g}_k^{\normalfont{\textrm{T}}}\mathbf{\hat{G}}^{*}\tilde{\boldsymbol{\lambda}}_l\rvert^2+\sigma_w^2}.\label{SINR when decoding common message MMSE approach}
 \end{equation}\normalsize

The SINR when decoding the private message at the $k$-th user is
\par\noindent\small
\begin{equation}
    \gamma_k=\frac{a_k^2\lvert \hat{\boldsymbol{g}}_k^{\normalfont{\textrm{T}}}\mathbf{\hat{G}}^{*}\tilde{\boldsymbol{\lambda}}_k\rvert^2}{d_k^{\left(v\right)}+\sum\limits_{\substack{j=1\\j\neq i}}^{N_c} a_{c_j}^2\lvert\left(\hat{\boldsymbol{g}}_k^{\normalfont{\textrm{T}}}+\tilde{\boldsymbol{g}}^{\normalfont{\textrm{T}}}_k\right)\mathbf{v}_{1}^{\left(j\right)}\rvert^2+\sum\limits_{\substack{l=1\\l\neq k}}^{K}\lvert\boldsymbol{g}_k^{\normalfont{\textrm{T}}}\mathbf{\hat{G}}^{*}\boldsymbol{\tilde{\lambda}}_l\rvert^2+\sigma_w^2}.\label{SINR when decoding private message MMSE approach}
\end{equation}\normalsize
 \end{proposition}
 \begin{IEEEproof}
 The average power of the received signal is
\par\noindent 
 \begin{align}
  &\mathbb{E}\left[\lvert y_k\rvert^2\right]=a_{c_i}^2 T_1+\sum\limits_{\substack{j=1\\j\neq i}}^{N_c} a_{c_j}^2\lvert\left(\hat{\boldsymbol{g}}_k^{\text{T}}+\tilde{\boldsymbol{g}}^{\text{T}}_k\right)\mathbf{v}_{1}^{\left(j\right)}\rvert^2+\sum_{l=1}^{K}a_l^2\lvert \boldsymbol{g}_k^{\text{T}}\mathbf{\hat{G}}^{*}\tilde{\boldsymbol{\lambda}}_l\rvert^2+\sigma_w^2,\label{Average power of the received signal MMSE approach}
 \end{align}
  where $\tilde{\boldsymbol{\Lambda}}=\left(\hat{\mathbf{G}}^{\text{T}}\hat{\mathbf{G}}^*+\frac{K\sigma_w^2}{P_t}\mathbf{I}_{K}\right)^{-1}$, whose $l$-th column is the vector $\tilde{\boldsymbol{\lambda}}_l$. From \eqref{Average power of the received signal MMSE approach}, we obtain \eqref{SINR when decoding private message MMSE approach} and \eqref{SINR when decoding common message MMSE approach}.
 \end{IEEEproof}

 The following proposition~\ref{prop:sinr_MMSE_RD} provides the SINR of the RU-MMSE-RD.
 \begin{proposition}
 \label{prop:sinr_MMSE_RD}
  For the RU-MMSE-RD, the SINR when decoding the common message at the $k$-th user in $\mathcal{K}_i$ is\par\noindent\small
 \begin{equation}
     \gamma_{c_i,k}=\frac{a_{c_i}^2\psi_{1,1}^{\left(i\right)^2}\lvert u_{k,1}^{\left(i\right)}\rvert^2}{d_{c_i,k}^{\left(v\right)}+\sum\limits_{\substack{j=1,j\neq i\\r\in \mathcal{K}_j}}^{N_c} a_r^2\lvert\hat{\boldsymbol{g}}^{\normalfont{\text{T}}}_k\hat{\mathbf{G}}^*_j\boldsymbol{\tilde{\lambda}}_{r\rightarrow q}^{\left(\textrm{j}\right)}\rvert^2 +\sum\limits_{\substack{l=1\\l\in\mathcal{K}_t}}^{K}a_l^2\lvert \boldsymbol{g}_k^{\normalfont{\text{T}}}\mathbf{\hat{G}}_t^{*}\boldsymbol{\tilde{\lambda}}^{\left(\textrm{t}\right)}_{l\rightarrow v}\rvert^2+\sigma_w^2/\varepsilon^2}.\label{SINR when decoding common message MMSE-RD approach}
 \end{equation}\normalsize
The SINR when decoding the private message at the $k$-th user is \par\noindent\small
\begin{equation}
    \gamma_k=\frac{a_k^2\lvert \hat{\boldsymbol{g}}^{\text{T}}_k\hat{\mathbf{G}}^*_i\boldsymbol{\tilde{\lambda}}_{k\rightarrow q}^{\left(\textrm{i}\right)}\rvert^2}{\sum\limits_{\substack{j=1,j\neq i\\r\in \mathcal{K}_j}}^{N_c} a_r^2\lvert\hat{\boldsymbol{g}}^{\normalfont{\textrm{T}}}_k\hat{\mathbf{G}}^*_j\boldsymbol{\tilde{\lambda}}_{r\rightarrow q}^{\left(\textrm{j}\right)}\rvert^2+\sum\limits_{\substack{l=1,l\neq k\\l \in \mathcal{K}_t}}^{K}\lvert\tilde{\boldsymbol{g}}_k^{\normalfont{\textrm{T}}}\mathbf{\hat{G}}^{*}_t\boldsymbol{\tilde{\lambda}}^{\left(t\right)}_{l\rightarrow v}\rvert^2+\sigma_w^2/\varepsilon^2}.\label{SINR when decoding private message MMSE-RD approach}
\end{equation}\normalsize
 \end{proposition}
 \begin{IEEEproof}
 The power of the received signal at user $k$ is
 \begin{align}
     \mathbb{E}\left[\lvert y_k\rvert^2\right]=&\varepsilon^2\left(a_{c_i}^2\psi_{1,1}^{\left(i\right)^2}\lvert u_{k,1}^{\left(i\right)}\rvert^2+d_{c_i,k}^{\left(v\right)}+\sum\limits_{\substack{j=1,j\neq i\\r\in \mathcal{K}_j}}^{N_c} a_r^2\lvert\hat{\mathbf{g}}^{\text{T}}_k\hat{\mathbf{G}}^*_j\boldsymbol{\tilde{\lambda}}_{r\rightarrow q}^{\left(\textrm{j}\right)}\rvert^2+\sum\limits_{\substack{l=1\\l\in\mathcal{K}_t}}^{K}a_l^2\lvert \mathbf{g}_k^{\text{T}}\mathbf{\hat{G}}_t^{*}\boldsymbol{\tilde{\lambda}}^{\left(\textrm{t}\right)}_{l\rightarrow v}\rvert^2\right)+\sigma_w^2.\label{received signal power RU-MMSE-RD}
 \end{align}
 From \eqref{received signal power RU-MMSE-RD}, we obtain \eqref{SINR when decoding common message MMSE-RD approach} and \eqref{SINR when decoding private message MMSE-RD approach}.
 \end{IEEEproof}
\vspace{-1.6em}
\subsection{Scaling of complexity and signaling}
\begin{figure}[t]
\begin{center}
\includegraphics[width=0.55\columnwidth]{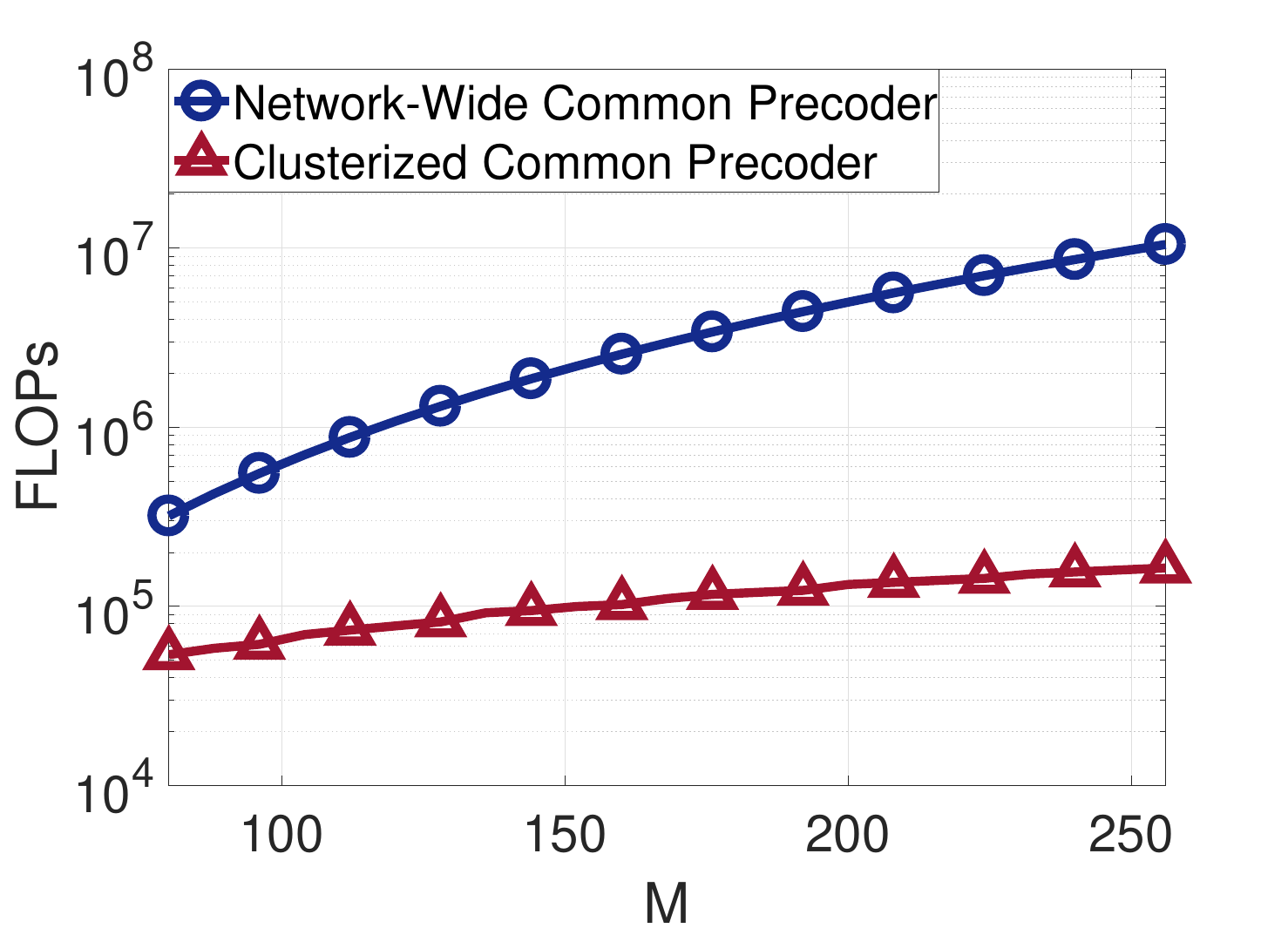}
\vspace{-1.5em}
\caption{Computational complexity of the common precoder in terms of  floating point operations per second (FLOPS), $N_c=M/32$.}
\label{FigComp}
\vspace{-2em}
\end{center}
\end{figure}
The proposed cluster-based precoders reduce the computational complexity of the system dramatically. For instance, Fig. \ref{FigComp} shows that the computational complexity of implementing a conventional common precoder is much higher than the computational complexity required by the proposed cluster-based approach. Furthermore, the following propositions state that the computational complexity of the cluster-based precoders remains bounded with the increase of the network size and the signaling load. We make the following Claim~\ref{clm:complex} and \ref{clm:scale}. 
\begin{claim}
\label{clm:complex}
Complexity scalability is guaranteed for the proposed cluster-based precoders.
\end{claim}
\begin{IEEEproof}
We use the number of complex multiplications and additions given by $M_c$ to describe the computational complexity. Given $K,N\to \infty$ 
with the ratio $K/N$ fixed, then $M_c/N$ has to be $\mathcal{O}\left(1\right)$ to ensure that the cost per AP remains fixed as the network grows in size.
For the cluster-based precoders, we have that $\mathcal{K}_i$ and $\mathcal{A}_k$  do not grow with $K$ and $N$. Then,
\begin{equation}
    \frac{M_c}{N}=\frac{\sum_{i=1}^{K} c\left(\lvert\mathcal{K}_i\rvert\right)}{N},\label{Scalability Number of Multiplications}
\end{equation}
where $c\left( Q \right)$ is the cost of inverting a square matrix of size $Q$. Since the cost in \eqref{Scalability Number of Multiplications} is $\mathcal{O}\left(1\right)$, its scalability is guaranteed.
\end{IEEEproof}

\begin{claim}
\label{clm:scale}
The proposed cluster-based precoders guarantee the scalability of the signaling required by the system.
\end{claim}
\begin{IEEEproof}
The number of channel estimates $L$ is determined by
\begin{equation}
    \frac{L}{N}=\frac{\sum_{i=1}^N\lvert\mathcal{A}_i\rvert}{N},
\end{equation}
which is $\mathcal{O}\left(1\right)$ and, therefore, scalable.
\end{IEEEproof}

\begin{figure}[t]
\begin{center}
\includegraphics[width=0.7\columnwidth]{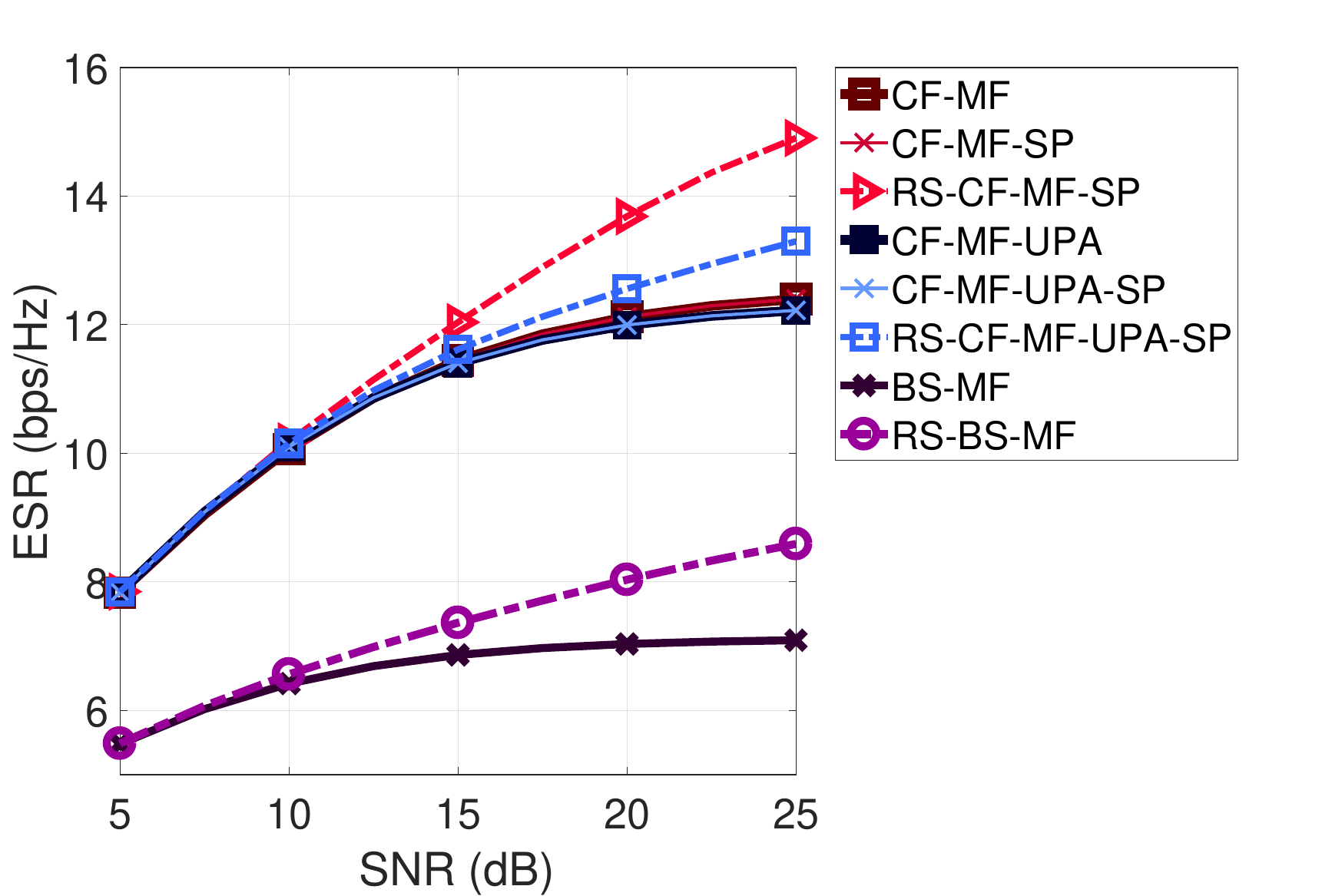}
\vspace{-1.5em}
\caption{Ergodic sum-rate performance of the proposed MIMO RS-CF with MF precoder under imperfect CSIT.}
\label{FigC3}
\vspace{-2em}
\end{center}
\end{figure}
\section{Numerical Experiments}
\label{sec:numexp}

We assess the performance of the proposed RS-CF scheme, cluster-based precoding, and power allocation algorithms against conventional CF systems and RS for cellular systems through numerical experiments. The notations "BS" and "RS-BS" in the legends denote, respectively, the conventional MIMO and the RS approach for systems based on cells with a centralized base station (BS). On the other hand, {we denote the conventional cell-free deployment and our proposed architecture by "CF" and "RS-CF", respectively}. The legends "SP" and "RD" stand for the proposed cluster-based sparse precoder and the proposed cluster-based precoder with reduced dimensions, respectively.



In all examples, we consider eight APs, four users, and imperfect CSIT, where the variance of the error is equal to $0.025$. The grid used to perform the power allocation has a step size of $0.05$. To compute the ASR  we employ $100$ error matrices $\tilde{\mathbf{G}}$. Furthermore, a total of $100$ channel realizations were considered to obtain the ESR, resulting in  $10000$ trials. The large-scale fading coefficients were defined as
 \begin{equation}
     \zeta_{m,k}=P_{m,k}\cdot 10^{\frac{\sigma^{\left(\textrm{s}\right)}z_{m,k}}{10}},
 \end{equation}
 where $P_{m,k}$ represents the path loss. The log-normal shadowing is modeled by  $10^{\frac{\sigma^{\left(\textrm{s}\right)}z_{m,k}}{10}}$, where $\sigma^{\left(\textrm{s}\right)}=8$ dB is the standard deviation and the random variable $z_{m,k}$ is Gaussian distributed with zero mean and unit variance. The path loss in dB is calculated using a three-slope model as\par\noindent\small
 \begin{align}
     P_{m,k}=\begin{cases}
  -L-35\log_{10}\left(d_{m,k}\right), & \text{$d_{m,k}>d_1$} \\
  -L-15\log_{10}\left(d_1\right)-20\log_{10}\left(d_{m,k}\right), & \text{$d_0< d_{m,k}\leq d_1$}\\
    -L-15\log_{10}\left(d_1\right)-20\log_{10}\left(d_0\right), & \text{otherwise,}
\end{cases}
 \end{align}\normalsize
 where $d_{m,k}$ is the distance between the $m$-th AP and $k$-th users, $d_1=50$ m, $d_0= 10$ m, and the attenuation $L$ is \par\noindent\small
 \begin{align}
L=&46.3+33.9\log_{10}\left(f\right)-13.82\log_{10}\left(h_{\textrm{AP}}\right)-\left(1.1\log_{10}\left(f\right)-0.7\right)h_u+\left(1.56\log_{10}\left(f\right)-0.8\right),
 \end{align}\normalsize
 where $h_{\textrm{AP}}=15$ m and $h_{u}=1.65$ are the positions of, respectively, the AP and the user equipment above the ground and frequency $f= 1900$ MHz.

 The noise variance is $\sigma_w^2=T_o k_B B N_f,$ where $T_o=290$ K is the noise temperature, $k_B=1.381\times 10^{-23}$ J/K is the Boltzmann constant, $B=20$ MHz is the bandwidth and $N_f=9$ dB is the noise figure. The SNR is defined as
 \par\noindent\small
 \begin{equation}
\text{SNR}=\frac{P_{t}\textrm{Tr}\left(\mathbf{G}^{\text{T}}\mathbf{G}^{*}\right)}{M K \sigma_w^2}.
 \end{equation}\normalsize

\begin{figure}
    \centering
  \subfloat[\label{4a}]{%
       \includegraphics[width=0.5\linewidth]{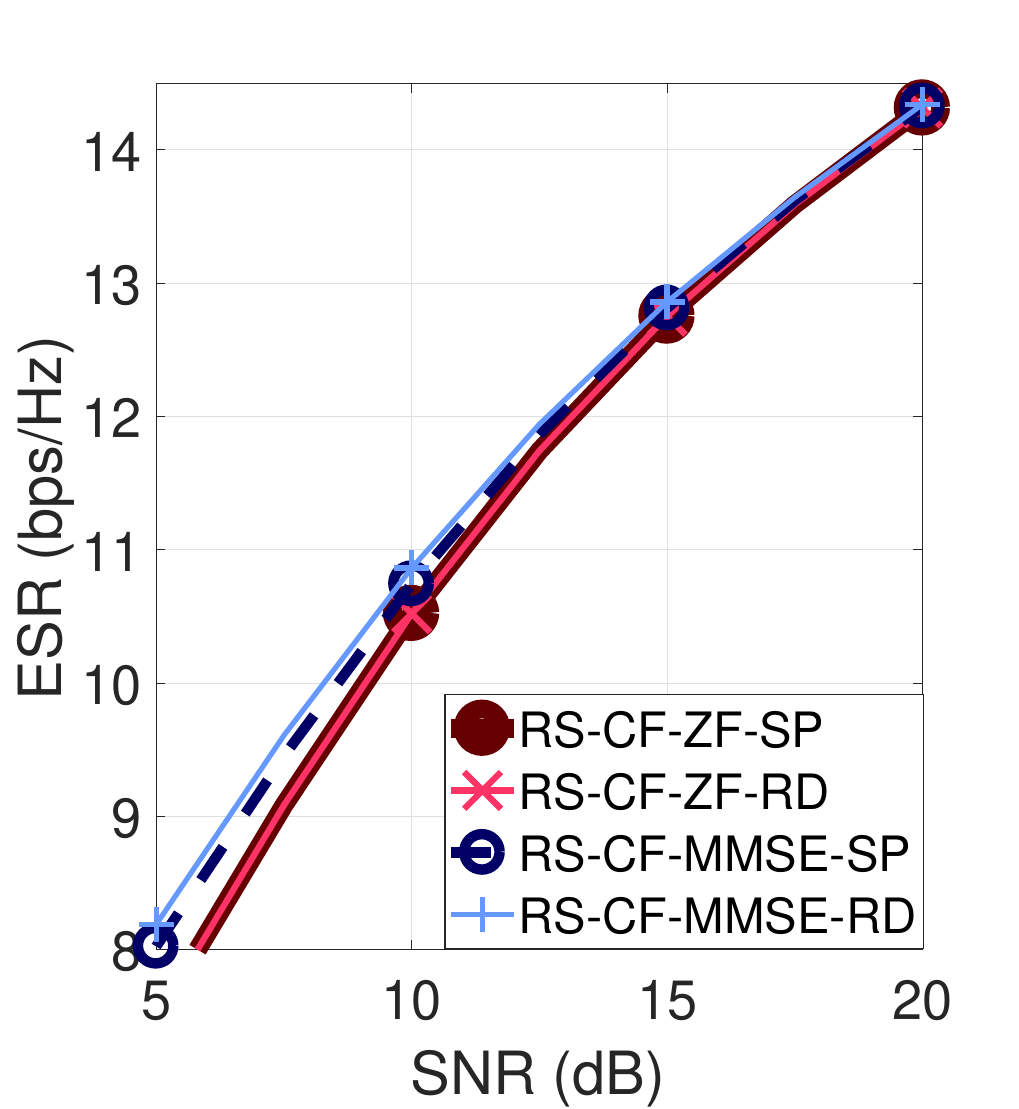}}
    \hfill
  \subfloat[\label{4b}]{%
        \includegraphics[width=0.5\linewidth]{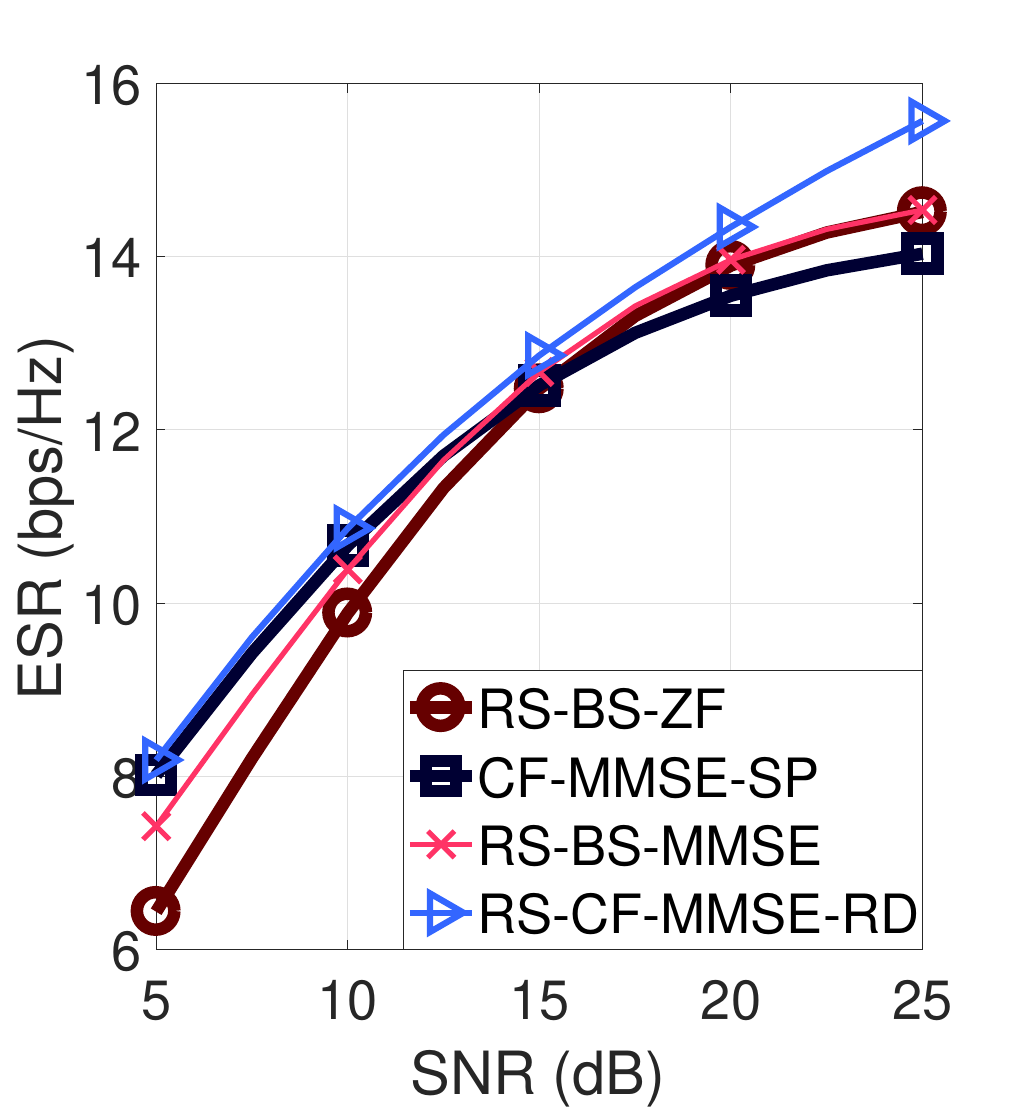}}
\vspace{-.75em}
  \caption{(a) ESR performance of the proposed MIMO RS-CF with SP and RD precoders. (b) ESR performance of the proposed MIMO RS-CF with ZF and MMSE precoders.}
  \label{FigC4}
\end{figure}

\begin{figure}
    \centering
  \subfloat[\label{5a}]{%
       \includegraphics[width=0.4\linewidth]{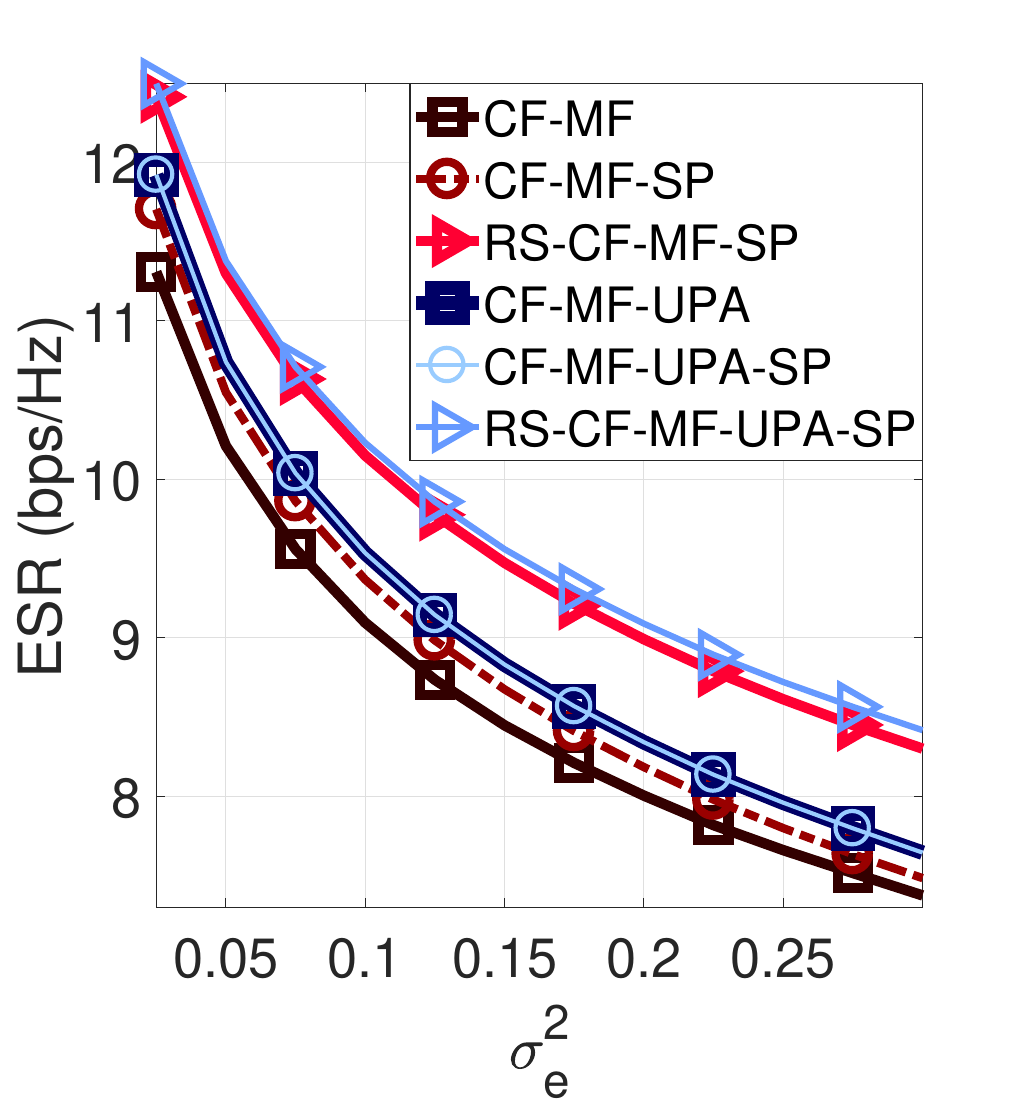}}
    \hfill
  \subfloat[\label{5b}]{%
        \includegraphics[width=0.4\linewidth]{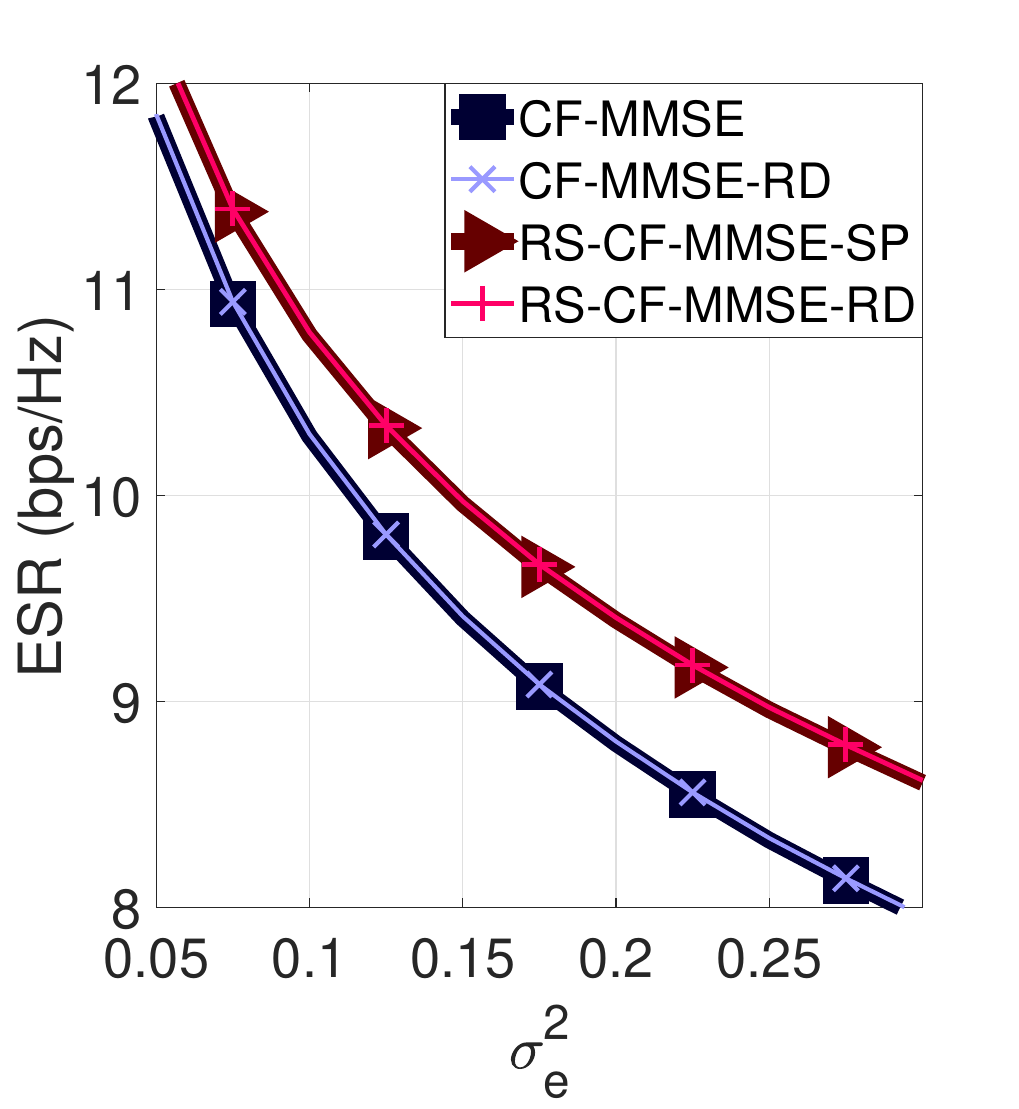}}
\vspace{-.75em}
  \caption{(a) ESR performance of the proposed MIMO RS-CF with MF precoder considering different qualities of CSIT. (b) ESR performance of the proposed MIMO RS-CF with MMSE precoder considering different qualities of CSIT.}
  \label{FigC5}
\end{figure}


 We first compare the ESR of the MF-precoder-based proposed RS-CF techniques against the conventional network-wide and sparse MF for CF systems, as shown in Fig. \ref{FigC3}. In the legend, the term ``UPA'' denotes uniform power allocation across private streams. {We observe that
 CF-based techniques outperform the conventional-BS-based systems thereby demonstrating the benefits of employing a distributed AP architecture. Moreover, the sparse precoder CF-MF-SP yields the same ESR as the network-wide CF-MF across all SNR levels. Similarly, the sparse CF-MF-UPA-SP and the network-wide CF-MF-UPA exhibit similar ESRs. In other words, the performance of sparse MF precoders does not deteriorate significantly when compared to the full matrix design.} This follows because the sparse matrix avoids channel coefficients associated with poor wireless links thereby saving power for better links. {More importantly,} the proposed RS-CF-MF-SP and RS-CF-MF-UPA-SP schemes outperform the conventional CF-MF scheme, which employs a network-wide precoder, and obtains an ESR gain of up to 10\%. Moreover, the proposed RS-CF architecture {also} outperforms the cell-based RS-BS-MF and BS-MF systems. {In summary, the proposed RS-CF is superior to both conventional CF and RS techniques for the BS.} This behavior was expected because the proposed cluster-based approach enhances the performance attained by the common streams through distributed APs and  transmission of multiple common streams.

In the second example whose results are shown in Fig. \ref{FigC4}, we assess the performance of the proposed ZF and MMSE precoders with the RS and RD techniques. In Fig. \ref{4a} we notice that the proposed {RS-CF-ZF-RD and RS-CF-MMSE-RD} precoders exhibit performance similar to the {sparse RS-CF-ZF-SP and RS-CF-MMSE-SP} precoders {, respectively.} {However, the inversion of matrices with reduced dimensions in the RD approach leads to lower computational complexity. Note that the RD performance depends on the design of the clusters of users. Fig. \ref{4a} shows that RD and SP yield similar ESR implying that the clusters have been properly designed. In contrast, a bad cluster design translates to additional interference thereby degrading the performance of the system.} The signaling load of {RD and SP} techniques is reduced because {both approaches employ a sparse channel matrix as a result of the cluster-based design.} In Fig \ref{4b}, we observe that the proposed RS-CF architecture with the proposed MMSE-RD precoder outperforms  conventional systems such as the CF with linear precoders and BS-based RS system. The proposed RS-CF scheme with cluster-based precoders exhibits an increasing ESR even in the high SNR regime, being more robust against CSIT imperfections than the CF and BS approaches. In contrast, the ESR of conventional schemes saturates because they cannot handle CSIT imperfections.
The best performance is exhibited by the RS-CF-MMSE-RD and the RS-CF-MMSE-SP schemes. Note that the performance of both algorithms is similar but the RD techniques have reduced complexity.


 Fig. \ref{5a} shows the ESR performance obtained  by the MF techniques considering different CSIT qualities at SNR of $20$ dB. {We deduce that the proposed RS-CF yields a significant gain over the conventional CF network. As shown in Fig. \ref{5a}, the  proposed RS-CF is also more robust than the conventional CF because the common stream allows the system to better deal with the interference originating from the CSIT imperfections. This} demonstrates the effectiveness of transmitting common streams {when compared to a CF system.} Similarly, Fig. \ref{5b} shows the performance of the MMSE precoder for different CSIT qualities. {It follows that CF-MMSE-RD and CF-MMSE-SP do not differ much in their performances. The proposed RS-CF-MMSE-RD and RS-CF-MMSE-SP also yield similar ESR plots.} Hence, again, the proposed RS-CF outperforms conventional CF deployments, {showing the effectiveness of employing common streams}. Moreover, RD techniques incur negligible performance loss as compared with the SP network-wide precoders, yet they reduce the computational complexity, which is crucial for practical systems. {The low performance loss in RD techniques is contingent on designing appropriate clusters.}

\begin{figure}[t]
\begin{center}
\includegraphics[width=0.45\columnwidth]{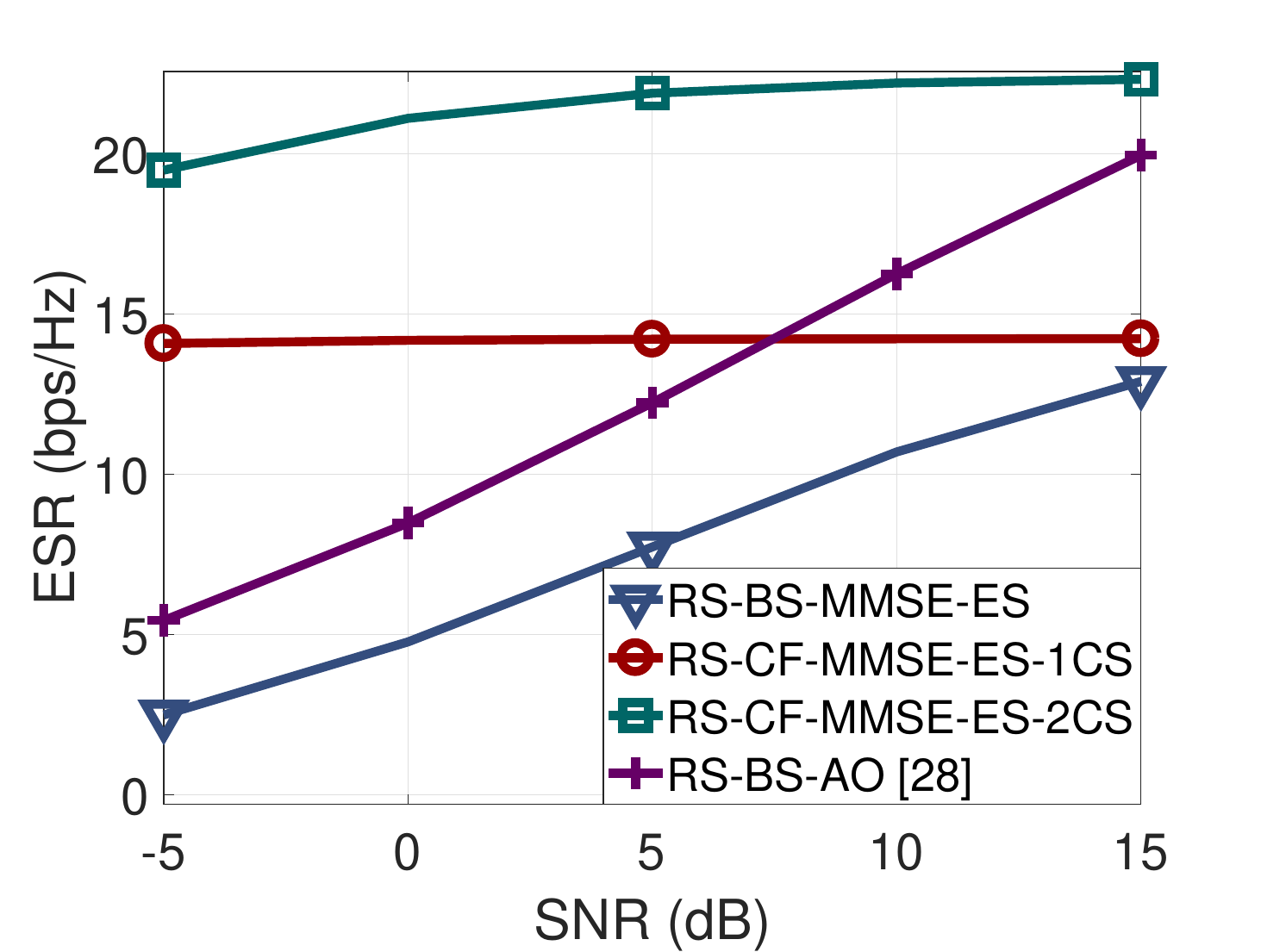}
\vspace{-1.5em}
\caption{ESR performance of the proposed MIMO RS-CF, $M=8$, $K=4$}
\label{FigAWMMSE}
\vspace{-2em}
\end{center}
\end{figure}

 In Fig. \ref{FigAWMMSE}, we compare the performance of the proposed MMSE-RD against the conventional alternating optimization (AO) algorithm proposed in \cite{JoudehClerckx2016} and used to solve the augmented weighted mean square error (AWMSE) problem. This AO algorithm maximizes the sum-rate by optimizing the precoders, the power allocation, and the receiver and is considered optimal for an RS-MISO system that employs only one common stream. The terms '1CS' and '2CS' denote the use of one or two common streams, respectively. For comparison, the power constraint is kept the same for the BS and CF approaches; so, both systems have the same transmit power. From the curves, we note that the ESR saturates for a single common stream. This behavior is expected because the performance of one common stream is limited by the performance of the worst user. In contrast, a substantial ESR gain is obtained by the proposed cluster-based RS-CF architecture employing two common streams. In this case, the performance of each common stream is limited by the worst user inside its cluster and not the worst user in the whole network. Therefore, the use of clusters {to transmit multiple common streams is an effective method to enhance the overall performance of CF-based systems.} {Even with the transmission of only two common streams, } our approach shows gains over the algorithm in \cite{JoudehClerckx2016}, which is optimal in terms of precoding and power allocation for a single common stream.

\begin{figure}[t]
\begin{center}
\includegraphics[width=0.4\columnwidth]{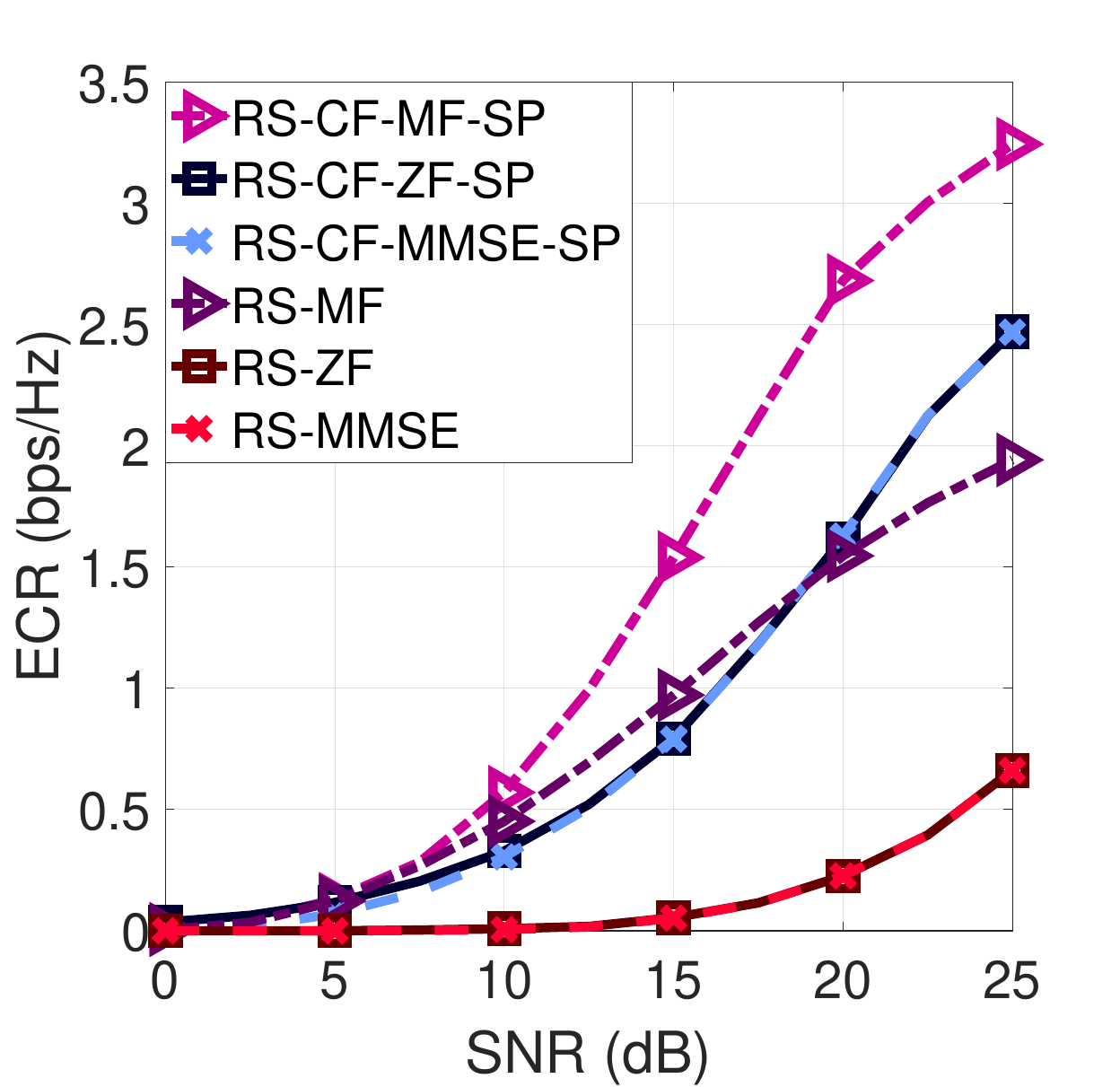}
\vspace{-1.5em}
\caption{Ergodic common rate performance of the proposed MIMO RS-CF, $M=8$, $K=4$, $N_c=2$.}
\label{FigC8}
\vspace{-2em}
\end{center}
\end{figure}

 In Fig. \ref{FigC8}, we examine the ECR performance of the proposed RS-CF scheme with SP precoders against the SNR and compare it with the conventional RS approach. The proposed RS-MMSE-SP and RS-ZF-SP schemes have a lower ECR performance than that of the RS-MF-SP scheme, whereas RS-CF outperforms the conventional RS approach. {This improvement follows from the fact that the distributed APs in the RS-CF architecture lead to better links for transmitting the common stream than the RS system. It is clear from Fig. \ref{FigC8} that the best ECR is obtained from the proposed RS-CF-MF-SP; its higher gain results from the MF precoder experiencing higher MUI levels than the other two linear precoders with the imcrease in SNR.} Therefore, RS allocates more power to the common stream to handle the interference efficiently, resulting in a higher ECR. We note that the RS-CF architecture yields a higher ECR than that of the conventional RS. The RS-CF architecture is fundamental to obtain this gain not only for the transmission of multiple common streams but also for the cluster-based approach, which limits the number of users that each common stream handles. The proposed Algorithm~\ref{alg:power} for power allocation with multiple common streams is also key to obtaining gains.
\vspace{-1em}

\section{Summary}
\label{sec:summ}
We have proposed a novel wireless communications architecture that combines the benefits of CF and RS techniques to yield a network that is more robust against imperfect CSIT. In particular, we developed cluster-based linear precoders that have low computational cost and signaling load. We obtained closed-form expressions to compute the sum-rate performance of the proposed RS-CF system equipped with such precoders. Our analysis of the proposed RS-CF system showed that it is robust against the imperfect CSIT and provides gains of up to $15 \%$ in ESR over previously suggested solutions. Numerical results support these observations by demonstrating that the RS-CF system with MF and ZF precoders outperforms existing RS-BS-MF, RS-BS-ZF, standard CF-MF, CF-ZF, BS-MF and BS-ZF systems in terms of the sum-rate.



\bibliographystyle{IEEEtran}
\bibliography{CellFree}

\end{document}